\documentclass[trackchanges, twocolumn]{aastex701}
\usepackage{comment}

\newcommand{\msun}{\mbox{M$_{\odot}$}}

\newcommand{\Aromal}[1]{{\color{red}[Aromal: #1]}}

\def\h2{$\rm H_2$}
\def\Nh2{$N$(H${_2}$)}

\def\Hb{\ensuremath{{\rm H}\beta}}
\def\kms{km\,s$^{-1}$}

\def\21{21-cm}

\def\t0{T$_{0}$}

\def\l{$\lambda$}

\def\c21{$C_{21}$}

\def\hii{H~{\sc ii}}

\def\U{$U_{336}$}
\def\B{$B_{438}$}
\def\V{$V_{606}$}
\def\Vm{$V_{547}$}
\def\I{$I_{814}$}
\def\Ha{$H_{\alpha_{665}}$}

\def\solarM{M$_\odot$}

\begin{document}

\title{Characterizing the star cluster populations in Stephan's Quintet using HST and JWST observations} 

\author[orcid=0009-0001-2178-4022,sname='Aromal',gname=Pathayappura]{P. Aromal}
\affiliation{Physics and Astronomy department, University of Western Ontario, 1151 Richmond Street, London, N6A 3K7, Ontario, Canada}
\affiliation{Institute for Earth and Space Exploration, Western University, 1151 Richmond St., London, ON N6A 3K7, Canada}
\email[show]{apathaya@uwo.ca}

\author[orcid=0000-0001-6217-8101,sname='Gallagher',gname=Sarah]{S. C. Gallagher}
\affiliation{Physics and Astronomy department, University of Western Ontario, 1151 Richmond Street, London, N6A 3K7, Ontario, Canada}
\affiliation{Institute for Earth and Space Exploration, Western University, 1151 Richmond St., London, ON N6A 3K7, Canada}
\email{sgalla4@uwo.ca}

\author[orcid=,sname=,gname=]{K. Fedotov}
\affiliation{Physics and Astronomy department, University of Western Ontario, 1151 Richmond Street, London, N6A 3K7, Ontario, Canada}
\affiliation{Institute for Earth and Space Exploration, Western University, 1151 Richmond St., London, ON N6A 3K7, Canada}
\email{}  

\author[orcid=0000-0001-5679-4215,sname=,gname=]{N. Bastian}
\affiliation{Astrophysics Research Institute, Liverpool John Moores University, Liverpool, L3 5RF, UK}
\email{}

\author[orcid=0000-0002-9471-5423,sname=,gname=]{U. Lisenfeld}
\affiliation{Departamento de Física Teórica y del Cosmos, Universidad de Granada, 18071 Granada, Spain}
\affiliation{Instituto Carlos I de Física Teórica y Computacional, Facultad de Ciencias, 18071 Granada, Spain}
\email{}  

\author[orcid=0000-0003-4877-9116,sname=,gname=]{J. C. Charlton}
\affiliation{Department of Astronomy and Astrophysics, The Pennsylvania State University, University Park, PA 16802, USA}
\email{}

\author[orcid=0000-0002-7607-8766,sname=,gname=]{P. N. Appleton}
\affiliation{Caltech/IPAC, MC 314-6, 1200 E. California Blvd., Pasadena, CA 91125, USA}
\email{}  

\author[orcid=0000-0003-1740-1284,sname=,gname=]{J. Braine}
\affiliation{Laboratoire d’Astrophysique de Bordeaux, Univ. Bordeaux, CNRS, B18N, allée Geoffroy Saint-Hilaire, 33615 Pessac, France}
\email{}  

\author[orcid=0000-0001-8348-2671,sname=,gname=]{K. E. Johnson}
\affiliation{Department of Astronomy, University of Virginia, Charlottesville, VA 22904, USA}
\email{}

\author[orcid=0000-0001-5737-5055,sname=,gname=]{P. Tzanavaris}
\affiliation{Center for Space Science and Technology, University of Maryland, Baltimore County, Baltimore, MD 21250, USA}
\affiliation{Center for Research and Exploration in Space Science and Technology, NASA Goddard Space Flight Center, Greenbelt, MD 20771, USA}
\affiliation{American Physical Society, Hauppauge, New York, NY 11788, USA}
\email{}

\author[orcid=0000-0003-2983-815X,sname=,gname=]{B. H. C. Emonts}
\affiliation{National Radio Astronomy Observatory, 520 Edgemont Road, Charlottesville, VA 22903, USA}
\email{} 

\author[orcid=0000-0001-5042-3421,sname=,gname=]{A. Togi}
\affiliation{Department of Physics, 601 University Dr, Texas State University, San Marcos, TX 78666, USA}
\email{}  

\author[orcid=0000-0002-1588-6700,sname=,gname=]{C. K. Xu}
\affiliation{Chinese Academy of Sciences South America Center for Astronomy, National Astronomical Observatories, CAS,
Beijing 100101, Peoples Republic of China}
\affiliation{National Astronomical Observatories, Chinese Academy of Sciences (NAOC), 20A Datun Road, Chaoyang District,
Beijing 100101, Peoples Republic of China}
\email{}  

\author[orcid=0000-0002-2421-1350,sname=,gname=]{P. Guillard}
\affiliation{Sorbonne Université, CNRS, UMR 7095, Institut d’Astrophysique de Paris, 98bis bd Arago, 75014 Paris, France}
\affiliation{Institut Universitaire de France, Ministère de l’Enseignement Supérieur et de la Recherche, 1 rue Descartes, 75231 Paris Cedex 05, France}
\email{}  

\author[orcid=0000-0003-0057-8892,sname=,gname=]{L. Barcos-Muñoz}
\affiliation{National Radio Astronomy Observatory, 520 Edgemont Road, Charlottesville, VA 22903, USA}
\email{}

\author[orcid=0000-0002-0806-168X,sname=,gname=]{L. J. Smith}
\affiliation{Space Telescope Science Institute, 3700 San Martin Drive, Baltimore, MD 21218, USA}
\email{}  

\author[orcid=,sname=,gname=]{I. S. Konstantopoulos}
\affiliation{Australian Astronomical Observatory, North Ryde, NSW 1670, Australia}
\email{}

\begin{abstract}

Stephan’s Quintet (SQ) is a local compact galaxy group system that exhibits significant star formation activity. A history of tidal interactions between its four member galaxies and a recent collision between an intruder galaxy and the original group are associated with active star formation, particularly in many shocked regions in the intra-group medium. 
Using an existing star cluster candidate (SCC) catalog constructed from HST UV/optical images, we integrate flux measurements from five near-infrared filters (F090W, F150W, F200W, F277W, F356W) obtained from JWST NIRCam observations in 2022. 
Leveraging the extended photometric baseline from HST and JWST, spanning $\sim$300 nm to $\sim$3500 nm, we perform spectral energy distribution (SED) fitting using the {\sc Cigale} code to derive reliable estimates of age, mass, and extinction for the 1,588 high-confidence SCCs.
We confirm earlier results that very young SCCs ($\sim$a few Myr) are predominantly located along previously identified shock regions near the merging galaxies, while older ($>100$~Myr) and  globular clusters are more widely distributed.  
Our analysis shows that NIR photometry helps break the age–extinction degeneracy, reclassifying many SCCs from older to younger, moderately dust-extincted clusters when added to HST-based SED fits.
We also observe a strong spatial correlation between young clusters and CO-traced molecular gas, although active star formation is present in several regions with no detectable CO. 
We find that the two prominent epochs of star formation, around 5 Myr and 200 Myr, correspond to the two major interaction events in SQ that gave rise to the observed extended tidal features.

\end{abstract}


\section{Introduction}

Compact groups are a class of dense galaxy environments with a relatively small number of galaxies (3 to tens) within a few galaxy radii of each other.  In contrast to galaxy clusters with similar number densities, compact group galaxies have the advantage of low relative velocities (a few 100s km~s$^{-1}$) that extend the time -- and therefore the impact -- of gravitational interactions from close encounters.  As a class, compact group galaxies show evidence for an unusual distribution of infrared spectral energy distributions (SEDs) that indicate that they rapidly transition from actively star-forming to quiescent \citep{Johnson+2007, Walker+2012}.  Compared to other environments (field and cluster), this characteristic SED distribution is most similar to what is seen in the infall region of galaxy clusters, where strong dynamical interactions first occur before galaxies have lost their cold gas \citep{Walker+2010}.  The infrared properties most directly tie to specific star formation rate \citep{Zucker+2016, Lenkic+2016} and provide strong evidence for accelerated galaxy evolution in compact group galaxies.  Furthermore, these systems might serve as local analogues for the early days of massive galaxy cluster formation.  
  
In gas-rich galaxies experiencing tidal interactions, star formation is often triggered, resulting in populations of young, massive star clusters.  As a result, epochs of star formation can be identified by searching for individual star clusters, identified as unresolved or marginally resolved point sources depending on the distance \citep[e.g.,][]{Whitmore1997}.   
As star clusters age, their colours become redder and luminosities fade \citep[e.g.,][]{bc03}.  With sufficient photometric wavelength coverage, the masses, ages, metallicities, and dust reddening of star clusters can be constrained by fitting the spectral energy distribution with single stellar population models \citep[e.g.,][]{Chandar+2010}. 
With this information in hand, star cluster populations have the potential to be traced to specific triggering events in the history of a compact galaxy group to provide  insight into the influence of the group environment on the evolution of its member galaxies. 
In compact galaxy groups, the small numbers of galaxies mean that the imprints of individual interactions last longer (because the galaxies have low relative velocities) and can lead to unusual scenarios.  For example, cold gas can be pulled out from a galaxy into the intragroup medium without getting heated to 10$^6$~K and becoming fully ionized. These gas clouds outside of galaxies are then available to participate in interactions in ways not often seen in the local universe.

Stephan's Quintet is one of the best-studied compact galaxy groups within 100~Mpc.  Originally identified by \citet{Stephan1877}, the group also met the selection criteria of \citet{hickson1982} and is catalogued as HCG~92.  Four of the galaxies in the quintet are physically close: NGC~7319, NGC~7318a, NGC~7318b, and NGC~7317, while NGC~7320 is an unrelated foreground system.  Dramatic distortions in the galaxy morphologies are evidence of a history of strong and complicated gravitational interactions.  Most strikingly, a $\sim50$~kpc-long strong shock that emits in radio continuum, X-rays, and 
strong emission from pure rotational mid-IR lines of H$_2$
through the intragroup medium \citep{Shostak1984, O'Sullivan2009, Appleton+2006, Cluver2010, appleton2023}.  This shock is the consequence of the relatively recent arrival of the gas-rich spiral galaxy, NGC~7318b (\mbox{$v_{\rm GSR}$ = 5967 km~s$^{-1}$}), which is understood to have moved from behind the group towards us at a velocity that is $\sim$700 km~s$^{-1}$ blueshifted from the mean group velocity.  From previous interactions, a galaxy's worth of H~{\sc i} ($1.35\times10^9$~\msun) and H$_2$ ($2.3\times10^8$~\msun) has been deposited into the intragroup medium at velocities from 6500--7000~km~s$^{-1}$ \citep[e.g.,][]{Williams2002,Lisenfeld2002}.  Bright H$\alpha$-emitting knots largely consistent with ongoing star formation (rather than being shock heated) are found in Stephan's Quintet A (SQA), a starburst to the North in the intragroup medium, along the shock, and in other structures primarily outside of the visible extent of the member galaxies \citep{puertas2021}.  

The star cluster population of Stephan's Quintet was first studied with high-resolution Hubble Space Telescope $BVI$ WFPC2 photometry by \citet{Gallagher2001}.  The colours of the  star clusters in the southern tidal tail of NGC~7319 were consistent with ages of $\sim150$~Myr, the age estimated by \citet{Moles1997} for the age of the feature, while the star cluster populations of SQA and the large shock ridge indicated much younger ages, $<10$~Myr.  More sensitive HST ACS observations supported and substantially extended these initial findings using more than 4 times as many clusters (496 compared to 115) found over a larger field of view.  Subsequent HST photometry added $U$ and medium-width $V$ band imaging centered on the main shock.  The choice of additional filters was made to break the age-reddening degeneracy for certain age ranges with only $BVI$ coverage \citep{fedotov_thesis}.  These studies established the value of high spatial resolution, precision photometry for constraining the epochs of star formation triggered by complex gravitational interactions in gas-rich environments.  

Recently, it has been shown that the addition of near-infrared photometry has the potential to greatly enhance the study of star clusters in nearby galaxies \citep{whitmore2023,rodriguez2023, sun2024, linden2024, levy2024}.
There are many studies in the literature where young embedded clusters, often missed in UV/optical images, are detected in JWST NIRCam images \citep{linden2023}.
Also, the additional photometric baseline in addition to UV/optical is useful for deriving better star cluster properties using SED-fitting routines.
We aim to achieve this by combining both HST and JWST NIRcam photometry to create a broad wavelength baseline that includes 10 photometric bands from NUV (HST \U\ band) to NIR (NIRCam F356W).
Accurate estimation of star cluster properties such as age, mass and dust extinction is not only important for the understanding of star cluster formation, but also shines light on the history of group interactions and how it led to star formation in the intragroup medium.

The paper is organized as follows. In Section \ref{sec:obs}, we provide the details of the observations, the sample selection and the {\sc Cigale} SED-fitting methods.
Section \ref{sec:results} presents our results based on derived star cluster properties from SED fitting.
Section \ref{sec:discussion} presents discussions based on our results.
In Section~\ref{sec:conclusions}, we summarize the main results of our work.
We assume a distance to Stephan’s Quintet of 94 Mpc and the
redshift is z = 0.0215 \citep{hickson1982}.
Throughout this paper, we use the flat $\Lambda$CDM cosmology with  $H_0$ = 70 \kms\ Mpc$^{-1}$ and $\Omega_{m,0}$ = 0.3.

\section{Observations $\&$  Cluster candidate selection}
\label{sec:obs}

\subsection{Data}

To study the star cluster population in Stephan’s Quintet, we make use of multi-wavelength imaging data from both the James Webb Space Telescope (JWST) and the Hubble Space Telescope (HST).
JWST observed the Stephan's Quintet field in June 2022.
The NIRCam and MIRI Early Release Observations of Stephan's Quintet (PID 2732) are described in \citet{pontoppidan2022}. The NIRCAM images were captured using six filters including F090W ($\lambda=0.9\mu$m), F150W($\lambda=1.5\mu$m) and F200W($\lambda=2\mu$m) in the short wavelength channel and F277W($\lambda=2.7\mu$m), F356W($\lambda=3.5\mu$m), and F444W($\lambda=4.4\mu$m) in the long wavelength channel. The total exposure duration in each filter is 2362 seconds.
Using the FULLBOX five-point ``5TIGHT'' dither pattern, three pairs of dithered tiles were used to create a rectangular mosaic that measured 6.3 × 7.3 arcmin$^2$. 
The level 2b images were obtained from the Mikulski Archive for Space Telescopes (MAST) at STScI in June 2024. 
All of the HST and JWST data presented in this paper can be accessed via\dataset[doi:10.17909/nqh1-x498]{http://dx.doi.org/10.17909/nqh1-x498}.

For UV/optical images, we used observations obtained with the HST Early Release Science data (ERS; proposal 11502, PI K. S. Noll). 
The observations were carried out with WFC3 in six filters of which we use F438W (\B), F606W (\V), F814W (\I), and F665N (\Ha) images. 
The exposure times were 13200, 5400, 7200, and 15600 seconds for the \B, \V, \I, and \Ha\ filters, respectively. 
Additionally, we added observations in the F336W (\U) and F547M
(\Vm) filters, which were obtained in October 2011 (proposal 12301, PI S. C. Gallagher), and
had exposure times of 17280 and 5834 seconds, respectively. 
The F336W filter images are essential to resolve age ambiguities at young ages, by completing the {\it UBVV$_m$IH$_{\alpha}$} photometric baseline \citep{Anders2004}.
The medium-band imaging with the F547M filter provides images that are free from nebular contamination by \Hb, [O III]$\lambda\lambda$4959, 5007, H${\alpha}$, [N II]$\lambda$6583 and [S II]$\lambda\lambda$6716, 6731, that is present in the very broad F606W filter images. 
This contamination was a limiting factor to our BVI age-dating process in a previous study of SQ \citep{fedotov2011} that used the V magnitudes from the F606W filter images.  The data analysis based on these images is presented in detail in \citet{fedotov2015} and briefly described in the next paragraph for completeness.

The twelve WFC3 images for each pointing of the \U, \B, \V, \Ha, and \I\ filters, and the four images for each pointing in the \Vm\ filter were combined into deep, cosmic-ray cleaned, geometric-distortion corrected mosaics using Multidrizzle version 3.3.8 \citep{fedotov_thesis}. 
Small shifts between images are caused by telescope pointing errors ($\leq 1\arcsec$). 
To correct these shifts, we measured the centroids of non-saturated point sources (e.g., stars in the broad band filters, as well as galaxy nuclei in \Ha) in each of the pointings and then computed the residual offsets between frames after an initial run of Multidrizzle. These offsets were then resupplied to the software in the form of a shift file to account for the offsets. This ensures the shape of the PSF is well-behaved and its width is minimized in the final output image.
Following the procedure outlined in \citet{fruchter2009}, we changed the values of the \textit{final pixfrac} and \textit{final scale} parameters to achieve consistent spatial resolution. The mosaics each have a pixel scale of 0$\arcsec$.028 pixel$^{-1}$. To correct the astrometry of the final mosaic products, we registered the images with the USNO B-1.0 all-sky astrometric catalog. 
This catalog was chosen for its completeness ($\sim$10$^9$ objects down to V = 21 mag) and an astrometric accuracy of 0$\arcsec$.2. 
Corrections to the world coordinate system (WCS) were made using the IMWCS task in WCSTools version 3.8.1.

\begin{deluxetable}{ccccccc}
\tablecaption{Details of HST and JWST observations of SQ used in this study \label{tab_obs}}
\tablecolumns{7}
\tablewidth{0pt}
\tablehead{
\colhead{Telescope} & \colhead{Filters} & \colhead{Bandwidth} & \colhead{PSF FWHM} & \colhead{Exposure time} & \colhead{Zero-point\tablenotemark{a}} & \colhead{$A_{mag}$\tablenotemark{a}} \\
\colhead{} & \colhead{} & \colhead{($\Delta \lambda$, nm)} & \colhead{(arcsec)} & \colhead{(s)} & \colhead{(Vega)}  & \colhead{(mag)}}
\startdata
    & \U\ & 55 & 0.075 & 17280  & 23.48 & 0.353  \\
    & \B\ & 67 & 0.070 & 13200  & 24.97 & 0.288  \\
HST & \Vm\ & 71 & 0.067  & 5834  & 24.75 & 0.226  \\
    & \V\ & 230 & 0.067 & 5400  & 25.99 & 0.197  \\
    & \I\ & 209 & 0.074 & 7200  & 24.68 & 0.122  \\
\hline  
     & F090W & 208 & 0.033 & 2362  & 26.24 & 0.096  \\
     & F150W & 335 & 0.050 & 2362  & 25.88 & 0.042  \\
JWST & F200W & 470 & 0.066 & 2362  & 25.59 & 0.026  \\
     & F277W & 706 & 0.092 & 2362  & 24.94 & 0.014  \\
     & F356W & 832 & 0.116 & 2362  & 24.61 & 0.009  \\
\enddata
\tablenotetext{a}{Refer subsection \ref{subsec:data} for a detailed description.}
\end{deluxetable}

\subsection{Source detection and photometry}
\label{subsec:data}

\begin{figure}
    \centering
    \includegraphics[viewport=0 0 780 370, width=\textwidth,clip=true]{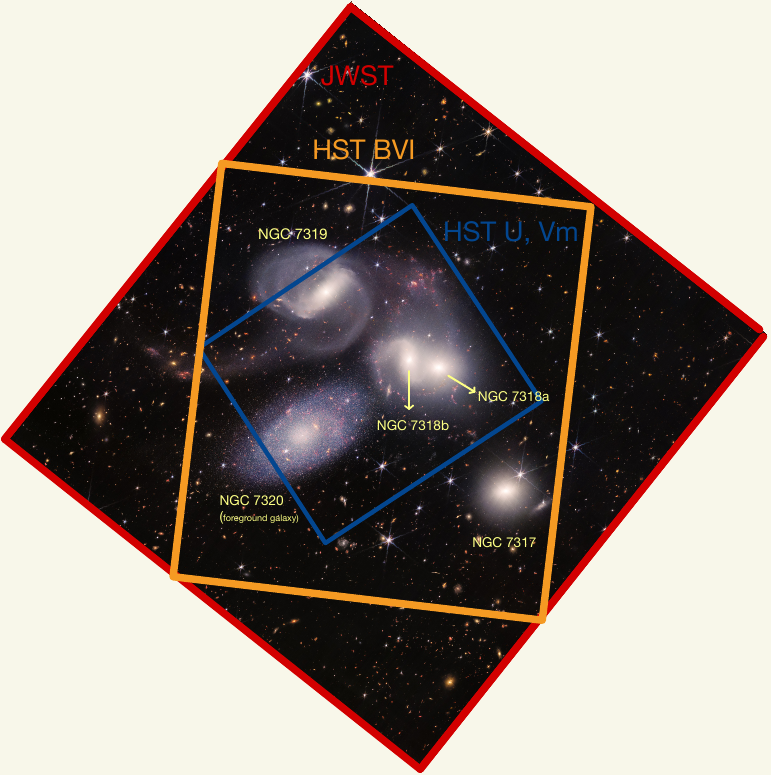}
    \caption{Left: The JWST NIRcam false color image (red square) is shown with the boxes inside representing the field of view for HST \B-\V-\I\ bands (orange) and HST \U-\Vm\ bands (blue). The five galaxies in the field are marked including the four galaxies in the group, namely NGC 7317, 7318a, 7318b, 7319 and the foreground galaxy NGC 7320.
    }
    \label{fig:hst_jwst_fov}
\end{figure}

With a distance of $\sim$ 94 Mpc to SQ \citep{hickson1982}, one pixel on the WFC3 image corresponds to 0$\arcsec$.028 ($\sim$ 12 pc). 
With the average size of a star cluster of 3--6 pc 
\citep{Barmby2006, Scheepmaker2007, Whitmore1997},
most of the star clusters in SQ will appear as point sources.
Even massive young clusters that can reach sizes of $\sim$ 10 pc 
(e.g., \citealt{Bastian2005a, Whitmore1999}; 
though both studies caution that the sizes larger than 10--15 pc are likely due to crowding) are not expected to be resolved. Therefore, our finding algorithm was tuned for point source detection in the HST images.
The detection of point sources was carried out with the DAOFIND task 
\citep{Stetson1987} in IRAF\footnote{NOIRLab IRAF is distributed by the Community Science and Data Center at NSF NOIRLab, which is managed by the Association of Universities for Research in Astronomy (AURA) under a cooperative agreement with the U.S. National Science Foundation.},
 with a 0.6$\sigma$ threshold in a median-divided 
 \V\ image
because it is the deepest HST image available. 
The detection process is detailed in \citet{fedotov2011} (Section 2.2 therein).
In order to maximize the photometric coverage, i.e, to complete the UBVV$_m$IH$_{\alpha}$ photometric baseline as mentioned before, we consider a smaller field of view for our analysis that overlaps between HST Early Release Science data (ERS; proposal 11502, PI K. S. Noll) and the later HST observations from 2011 (proposal 12301, PI S. C. Gallagher).
The latter observations in \U\ and \Vm\ bands had a smaller field of view compared to the earlier ERS HST observations in BVI bands as shown in Figure~\ref{fig:hst_jwst_fov}.

We used the list of coordinates of detected sources to derive the PSF magnitudes of these sources in each filter. 
The PSF was constructed from bright, isolated, and unsaturated stars with point-like radial curves of growth.
There were 18, 60, 27, 86, and 89 such stars in the \U, \B, \Vm, \V, and \I\ filters,
respectively.
PSF photometry was carried out with the DAOPHOT package \citep{Stetson1987} in IRAF, and local background levels were estimated from annuli with radii of 3–5 pixels for each source.
We converted the instrumental magnitudes to the VEGAMAG magnitude system by applying
the following zero points: 23.484 mag for \U, 24.974 mag for \B, 24.748 mag for \Vm,
25.987 mag for \V, and 24.680 mag for \I\ as provided by STScI.\footnote{Space Telescope Science Institute: http://www.stsci.edu/hst/wfc3/phot\_zp\_lbn}
The photometry in the F336W, F438W, F606W, and F814W wide filters was corrected for
foreground extinction calculated using data from 
\citet{Schlafly2011}, with resulting values of A$_{336}$ = 0.353 mag, A$_{438}$ = 0.288 mag, A$_{606}$ = 0.197 mag, and A$_{814}$ = 0.122 mag. 
The correction for the medium F547M filter was interpolated, based on the wide filter correction, to have a value of A$_{547}$ = 0.226 mag. 
The aperture corrections were determined in a similar manner as in \citet{fedotov2011}, and have the following values: 0.790, 0.724, 0.641, 0.600,
and 0.599, for \U, \B, \Vm, \V, and \I\ filters, respectively.
In total, $5996$ point sources were detected in all 5 HST filters (\U, \B, \Vm, \V, and \I) and constitute our initial, unfiltered star cluster candidate catalog. 
Note that we exclude the \Ha\ images from the construction of this catalog, as the narrow-band data were not sufficiently deep to achieve the sensitivity required for detecting low-luminosity cluster candidates when compared to the other wide and medium-band images used in this study.

Next, we extend the photometric coverage of this catalog obtained from HST photometry to infrared wavelengths using new JWST NIRCam observations \citep{pontoppidan2022}.
We use the source coordinates from the HST star cluster candidate catalog and carry out photometry in the JWST NIRCam images (F090W, F150W, F200W, F277W and F356W) at the same positions.
Note that in this paper we do not run independent point source detection in JWST images.
We measure a slight astrometric offset between the catalog based on HST images and the JWST images and therefore applied $\Delta$RA = 0.14\arcsec, and $\Delta$DEC = --0.18\arcsec\ corrections to the point-source catalog before performing the photometry on the JWST images.
We exclude F444W photometry from our analysis, as this filter has a larger PSF (FWHM $\sim$ 0.145\arcsec) than shorter wavelength filters and may include potential contamination from either a hot dust component or the circumstellar dust of AGB stars that could affect individual star cluster spectral energy distributions (SEDs) in the NIR \citep{linden2023}.

The JWST PSF photometry for the point source catalog was performed using the publicly available Python package Photutils \citep{larry_bradley_2024}. For this, NIRCam PSFs were generated using the EPSFBuilder class in Photutils, following the same methodology as for HST photometry.
Using these model PSFs, instrumental magnitudes were calculated with PSF photometry after locally subtracting the background for each source. The background estimation was carried out using the DAOPHOT MMM algorithm \citep{Stetson1987} as implemented in Photutils. We used annuli with radii of 3--5 pixels to compute the local background level around each source.
The VEGAMAG zeropoints used for NIRCam instrumental magnitudes are as follows :
26.24 mag for F090W, 25.88 mag for F150W, 25.59 mag for F200W, 24.94 mag for F277W and 24.61 mag for F356W.\footnote{https://jwst-docs.stsci.edu/jwst-near-infrared-camera/nircam-performance/nircam-absolute-flux-calibration-and-zeropoints}
The aperture corrections (10 pixel to infinity) were calculated using WebbPSF software\footnote{https://stpsf.readthedocs.io/en/latest/} \citep{perrin2014_webbpsf} which gives approximately 0.1 mag for all NIRCam filters. 
Following \citet{draine2003}, we also correct for the foreground extinction (A$_{mag}$) in the NIRCam filters.
Except for the F090W filter (A$_{0.9 \mu m} \sim 0.1$ mag), these corrections are smaller compared to HST filters and do not appear to affect magnitudes considerably. 
The details of observations are summarized in Table~\ref{tab_obs}.
In the next subsection, we look at how we filter the point source catalog to create a final star cluster candidate (SCC) catalog for our analysis.

\subsection{The selection criteria for star cluster candidates}
\label{subsec:sel_criteria}

From an unfiltered list of all 5996 detected point sources in the field, we created a catalog of high-confidence star cluster candidates (SCCs) using several selection criteria \citep{fedotov2011, fedotov_thesis}. 
It is important to note that the sample selection is entirely based on HST photometry and that the JWST photometry is used only for the purpose of extending the photometric baseline of this sample. 
The selection criteria used in \citet{fedotov_thesis} are summarized as follows:

\begin{itemize}
    \item Color cuts (S1): To minimize the contamination from the foreground stars, we keep only objects that satisfy the following color restrictions: \B--\V\ $<$ 1.5 or \V--\I\ $<$ 1.0. These values were chosen based on a spectroscopic study of sources in SQ by \citet{Trancho2012} (see Fig. 8 therein).

    \item Photometric error $\sigma$ $<$ 0.3 mag in all five filters (S2): For an object to be considered a SCC, its photometric errors should be less than 0.3 in all filters. This requirement reduces the uncertainty in cluster age and mass estimated subsequently.

    \item Sharpness between --2 and 2 in all bands (S3): 
    To further minimize contamination from cosmic rays and background galaxies, we applied a sharpness filter which is a constraint on the intrinsic angular size of the detected objects.
    Sharpness is measured as the difference between the square of the FWHM of the detected point source and the square of the FWHM of the PSF. 
    But we note that several extremely large star clusters and star cluster complexes can potentially be filtered out by the application of the sharpness criterion.

    \item Magnitude cut at M$_{V_{606}}$ $< -9$ mag (S4): To eliminate contamination from individual luminous supergiants, which can reach M$_{V_{606}}$ $\sim -8$ mag, we considered only sources with absolute magnitude M$_{V_{606}}$ $< -9$  mag.

    \item Goodness of fit $\chi^2$ in I-band (S5): The I-band was chosen for the DAOPHOT goodness of fit factor ($\chi^2$) criterion because the PSF model is best determined in that band and there is no expected contamination from emission lines. Thus, this band will give a more realistic result in the determination of the goodness of the PSF fitting. For this condition, we require $\chi^2$ $<$ 3.
\end{itemize}

After applying these criteria to the list of point sources, a final sample of 1,588 sources was obtained that satisfied all the conditions \citep{fedotov_thesis}. 
We refer to this as our high-confidence SCC sample and all subsequent analyses in this paper are based on this sample. 
In a future study, we plan to expand the sample by detecting additional point sources in JWST images, making use of its better sensitivity (Aromal et al, in preparation). However, this requires a more detailed analysis and is beyond the scope of this paper.

It is important to point out several limitations inherent to our selection criteria. First, the point source detection algorithm may fail to identify a significant number of SCCs in regions characterized by strong extended or diffuse emission, such as SQ-A and Left arc (see subsection~\ref{subsec:age_vs_mass}).
In addition, source detection is impacted by crowding in several regions, most notably in SQ-A, where a portion of the central clump is missed while point sources in the periphery are reliably recovered.
As a result, all integrated quantities derived from our analysis such as total SFR should be interpreted as robust lower limits. Second, the requirement that all sources be detected in the \U\ band likely biases our sample against highly extinct SCCs. 
This effect is particularly relevant for very young clusters that remain embedded within their natal molecular clouds, and may thus escape detection in the UV bands.

As mentioned above, we used the HST SCC catalog to perform PSF photometry on JWST NIRCam images.
We find that most SCCs are detected in JWST images as well.
The fraction of non-3 $\sigma$ detections in each JWST filter are 0.15, 0.14, 0.11, 0.14 and 0.17 for F090W, F150W, F200W, F277W and F356W respectively.
Only 3\% of the sources have non-detections in all the JWST filters.
Thus, we confirm that nearly all SCCs in our catalog have at least one JWST photometric measurement.
Furthermore, we confirm that the quality-of-fit metric—defined as the absolute value of the sum of the fit residuals divided by the fit flux—yields sufficiently good values for the NIRCam photometry. This reinforces our confidence in the reliability of the JWST photometry for the SCC sample.
Table~\ref{tab_hst} and Table~\ref{tab_jwst} provide the details of PSF photometry fluxes in HST and JWST filters of our high confidence SCC sample respectively; the full version of this table is available in machine-readable format in the online journal.

\begin{deluxetable}{ccccccc}
\tablecaption{PSF photometry of HST observations (in mJy)\tablenotemark{*} \label{tab_hst}}
\tablecolumns{7}
\tablewidth{0pt}
\tablehead{
\colhead{RA} & \colhead{DEC} & \colhead{F336W} & \colhead{F438W} & \colhead{F547M} & \colhead{F606W} & \colhead{F814W} \\
\colhead{} & \colhead{} &
\colhead{(10$^{-4}$ mJy)} & \colhead{(10$^{-4}$ mJy)} &
\colhead{(10$^{-4}$ mJy)} & \colhead{(10$^{-4}$ mJy)} & \colhead{(10$^{-4}$ mJy)}
}
\startdata
339.006260 & 33.961269 & $8.38 \pm 0.22$ & $9.92 \pm 0.57$ & $7.78 \pm 0.28$ & $8.94 \pm 0.25$ & $5.76 \pm 0.17$ \\
338.981840 & 33.962265 & $7.32 \pm 0.28$ & $8.57 \pm 0.42$ & $6.86 \pm 0.30$ & $9.07 \pm 0.37$ & $6.77 \pm 0.24$ \\
338.984920 & 33.963043 & $4.99 \pm 0.31$ & $6.83 \pm 0.37$ & $4.21 \pm 0.18$ & $5.95 \pm 0.27$ & $3.15 \pm 0.19$ \\
338.994290 & 33.962212 & $4.08 \pm 0.18$ & $4.70 \pm 0.19$ & $2.79 \pm 0.12$ & $5.68 \pm 0.29$ & $1.98 \pm 0.11$ \\
338.997740 & 33.960545 & $3.91 \pm 0.15$ & $9.47 \pm 0.37$ & $7.57 \pm 0.20$ & $9.27 \pm 0.26$ & $7.78 \pm 0.28$ \\
\enddata
\tablenotetext{*}{The full version of this table is available in machine-readable format in the online journal.}
\end{deluxetable}

\begin{deluxetable}{ccccccc}
\tablecaption{PSF photometry of JWST observations (in mJy)\tablenotemark{*} \label{tab_jwst}}
\tablecolumns{7}
\tablewidth{0pt}
\tablehead{
\colhead{RA} & \colhead{DEC} & \colhead{F090W} & \colhead{F150W} & \colhead{F200W} & \colhead{F277W} & \colhead{F356W} \\
\colhead{} & \colhead{} &
\colhead{(10$^{-4}$ mJy)} & \colhead{(10$^{-4}$ mJy)} &
\colhead{(10$^{-4}$ mJy)} & \colhead{(10$^{-4}$ mJy)} & \colhead{(10$^{-4}$ mJy)}}
\startdata
339.006260 & 33.961269 & $4.18 \pm 0.16$ & $1.92 \pm 0.07$ & $1.26 \pm 0.04$ & $0.92 \pm 0.05$ & $0.62 \pm 0.03$ \\
338.981840 & 33.962265 & -- & -- & $6.26 \pm 0.85$ & $5.04 \pm 0.37$ & $4.53 \pm 0.21$ \\
338.984920 & 33.963043 & $1.89 \pm 0.13$ & $1.92 \pm 0.13$ & $2.04 \pm 0.14$ & $1.98 \pm 0.20$ & $2.53 \pm 0.13$ \\
338.994290 & 33.962212 & $1.98 \pm 0.16$ & $0.72 \pm 0.06$ & $1.37 \pm 0.97$ & $1.47 \pm 1.50$ & $2.70 \pm 1.19$ \\
338.997740 & 33.960545 & $18.60 \pm 7.60$ & $5.67 \pm 0.23$ & $5.21 \pm 0.12$ & $3.95 \pm 0.13$ & $2.81 \pm 0.64$ \\
\enddata
\tablenotetext{*}{The full version of this table is available in machine-readable format in the online journal.}
\end{deluxetable}

\subsection{SED fitting using {\sc Cigale}}
\label{subsec:cigale}

For spectral energy distribution (SED) fitting analysis of star cluster candidate (SCC) sample, we use the photometry in the five bands in HST (\U, \B, \V, \Vm\ and \I) and the additional five bands in JWST NIRCam (F090W, F150W, F200W, F277W and F356W).
We perform the SED fitting using the publicly available package {\sc Cigale}: Code Investigating
GALaxy Emission \footnote{https://cigale.lam.fr/} 
to estimate various star cluster properties
\citep{cigale2019}.
Although {\sc Cigale} is commonly used for the SED fitting of galaxies/AGNs, \citet{turner2021} carried out benchmark analyzes to validate the code for star cluster SED fitting. Their study used a sample of identified star clusters in NGC 3351 as part of the Physics at High Angular Resolution in Nearby GalaxieS-HST (PHANGS-HST) survey. 
{\sc Cigale} follows the energy balance principle, ensuring that the energy absorbed by dust in the UV-optical is re-emitted in the infrared, which is crucial for obtaining reliable extinction and age estimates especially considering the inclusion of NIR photometry in our studies.

{\sc Cigale} works by generating a grid of SED models depending on the input parameters and comparing it with the available photometric data.
Here, we use two free parameters to sample the model grids: (1) age of the star cluster ($\tau$) and (2) dust reddening ($A_V$), based on single-age stellar population (SSP) models. The sampling grid is chosen as : (1) for age, we have linearly spaced grid in the range of 1-20 Myr ($\Delta \tau = 1$ Myr) and uniformly log-spaced grids from 20-10$^4$ Myr ($\Delta log(\tau, yr)$ = 0.05), 
(2) For reddening, we have linearly spaced grid of $A_V$ = 0-1.6 ($\Delta$ A$_V$ = 0.1).
The upper limit of $A_V$=1.6 is chosen for two key reasons:
(i) Balmer decrement measurements of H II regions in Stephan’s Quintet indicate low to moderate dust extinction with a median A$_V \sim 1$ \citep{konstant2014, puertas2021}. 
Furthermore, the fact that we detect these clusters in \U, where high extinction would severely suppress flux, further supports the assumption of moderate extinction.
(ii) SED fitting studies by \citet{whitmore2020} demonstrate that allowing large extinction values can sometimes lead to misidentification of older clusters as young ones. To mitigate this issue, they recommend constraining the range of $A_V$ values using independent methods such as the Balmer decrement.
By setting a moderate upper limit on $A_V$, we ensure more reliable age estimates while avoiding potential degeneracies in SED fitting.
Additionally, the cluster mass (m) is another important output parameter, but this is fixed by the age-reddening grid given an initial mass function (IMF) and star formation history (SFH).

We used \citet{ssp2003} as the SSP model with a solar metallicity value of 0.02 and the IMF was taken from \citet{chabrier2003}. 
For star formation history, we adopt the method mentioned in \citet{turner2021} to incorporate an instantaneous star-formation model in {\sc Cigale} as required in the case of star clusters. 
This is done by using the {\it{sfh2exp}} star-formation history (SFH) in {\sc Cigale} with a very short e-folding time ($= 10^{-5}$ Myr) to represent an instantaneous burst of star formation.
Importantly, we consider the contribution of nebular emission lines in our SED fitting to improve the accuracy of estimated parameters.
In addition, \citet{charlot2000} and \citet{dale2014} are chosen as models for dust attenuation and dust emission respectively in {\sc Cigale}.
The details of all the {\sc Cigale} input parameters are summarized in Table~\ref{tab:cigale_tab}.
An additional, optional input parameter, $\mu$, allows for the inclusion of dust attenuation from stellar birth clouds in addition to that from the ISM. While this parameter is excluded from our fiducial runs to avoid introducing additional uncertainties, its effects are considered and discussed in detail in subsection~\ref{subsec:input_vary}.

After the model grids are finalized, the observed photometric fluxes are provided as input to the code.
We conduct two {\sc Cigale} runs: (i) using only HST photometry and (ii) incorporating both HST and JWST photometry. This allows us to assess the impact of NIR photometry on the derived SED fitting parameters, as discussed in Section \ref{subsec:age_Av_degeneracy}.
{\sc Cigale} then performs the SED fitting by comparing the photometric flux in each filter with the SED models created based on the model parameters mentioned above.
The pdf$\_$analysis module in {\sc Cigale} then estimates the physical properties using likelihood-weighted parameters on the fixed model grid.
We carry out the SED fitting for all 
the 1,588 high-confidence SCCs considered in our sample.
The {\sc Cigale} output files include the best fit parameters from the input grids and also provide best fit SED spanning a large wavelength range of interest ($\sim$100-5000 nm).
In the next section, we present the results on the star cluster properties as obtained from {\sc Cigale}.

\begin{deluxetable}{cc}
\tablecaption{{\sc Cigale} Model parameters \label{tab:cigale_tab}}
\tablecolumns{2}
\tablewidth{0pt}
\tablehead{}
\startdata
    Initial Mass Function & \citet{chabrier2003} \\
    Star Formation History  & Instantaneous star formation  \\
    Age & 1-20 in 1 Myr steps, \\
    & 20-10$^4$ Myr in 0.05 dex steps\\
    Extinction, A$_V$ &  0-1.6 in 0.1 steps \\
    SSP model & \citet{ssp2003} \\
    Metallicity (Z) & 0.02 \tablenotemark{*} \\
    Dust attenuation model &  \citet{charlot2000} \\
    Dust emission model & \citet{dale2014} \\
    \hline
\enddata
\tablenotetext{*}{Z is defined in mass fraction of the metals in a star as considered in the SSP model.}
\end{deluxetable}

\begin{deluxetable*}{cc|ccc|cccc}
\tablecaption{{\sc Cigale} SED fitting results with HST+JWST photometry\tablenotemark{*} \label{tab_cigale}}
\tablecolumns{9}
\tablewidth{0pt}
\tablehead{
\colhead{RA} & \colhead{DEC} & 
\multicolumn{3}{c}{\textbf{Fiducial run}} & 
\multicolumn{4}{c}{\boldmath$\mu$ \textbf{(free parameter)\tablenotemark{a}}} \\
\colhead{} & \colhead{} & 
\colhead{log(age)} & \colhead{log(mass)} & \colhead{$A_{V}$} &
\colhead{log(age)} & \colhead{log(mass)} & \colhead{$A_{V, \mathrm{ISM}}$} & \colhead{$\mu$} \\
\colhead{} & \colhead{} & 
\colhead{(yr)} & \colhead{(\msun)} & \colhead{(mag)} & 
\colhead{(yr)} & \colhead{(\msun)} & \colhead{(mag)} & \colhead{}
}
\startdata
339.006260 & 33.961269 & 6.6990 & 4.0829 & 0.00 & 6.6990 & 4.0829 & 0.00 & -- \\
338.981840 & 33.962265 & 6.8451 & 4.4034 & 0.35 & 6.9031 & 4.3709 & 0.05 & 0.6 \\
338.984920 & 33.963043 & 6.6021 & 3.9567 & 0.35 & 6.6021 & 3.9567 & 0.35 & 1.0 \\
338.994290 & 33.962212 & 6.6021 & 3.8752 & 0.45 & 6.6021 & 3.8752 & 0.45 & 1.0 \\
338.997740 & 33.960545 & 8.0569 & 5.2516 & 0.00 & 8.0569 & 5.2516 & 0.00 & -- \\
338.998750 & 33.960556 & 6.8451 & 4.3157 & 0.60 & 6.8451 & 4.2802 & 0.20 & 0.4 \\
\enddata
\tablenotetext{a}{$\mu = \frac{A_{V, \mathrm{ISM}}}{A_{V, \mathrm{BC}} + A_{V, \mathrm{ISM}}}$ as defined in Section~\ref{subsubsec:mu}.}
\tablenotetext{*}{The full version of this table is available in machine-readable format in the online journal.}
\end{deluxetable*}

\begin{deluxetable}{ccccc}
\tablecaption{{\sc Cigale} SED fitting results with HST photometry only\tablenotemark{*} \label{tab_cigale_hst}}
\tablecolumns{5}
\tablewidth{0pt}
\tablehead{
\colhead{RA} & \colhead{DEC} & 
\colhead{log(age)} & \colhead{log(mass)} & \colhead{$A_{V}$} \\
\colhead{} & \colhead{} &
\colhead{(yr)} & \colhead{(\msun)} & \colhead{(mag)}
}
\startdata
339.006260 & 33.961269 & 6.6990 & 4.2677 & 0.45 \\
338.981840 & 33.962265 & 6.8451 & 4.3086 & 0.30 \\
338.984920 & 33.963043 & 6.6990 & 3.9521 & 0.25 \\
338.994290 & 33.962212 & 6.6021 & 3.7289 & 0.25 \\
338.997740 & 33.960545 & 8.0569 & 5.2159 & 0.00 \\
\enddata
\tablenotetext{*}{The full version of this table is available in machine-readable format in the online journal.}
\end{deluxetable}

\section{Results}
\label{sec:results}

In this section, we present the derived properties of SCCs from {\sc Cigale} SED fitting and analyze their distributions and spatial variations. We examine how the inclusion of JWST photometry, in addition to HST photometry, affects the parameter estimations. Furthermore, we analyze the relationships between different SCC properties based on the SED fits from HST+JWST photometry and how they vary according to the locations of the SCCs within the Stephan’s Quintet (SQ) field.

\subsection{The derived SCC properties from CIGALE runs}
\label{subsec:age_Av_degeneracy}

The distribution of derived star cluster properties, including cluster age ($\tau$), mass (m) and dust extinction ($A_V$) as obtained from {\sc Cigale} runs is shown in Figure~\ref{fig:hst_jwst_comp}. 
The best-fit parameter values from the {\sc Cigale} SED fits with HST+JWST photometry and HST photometry only are provided in Table~\ref{tab_cigale} and Table~\ref{tab_cigale_hst} respectively.
One of the key results of our analysis from HST and HST+JWST runs is that the cluster properties remain largely consistent for the majority of SCCs in our sample.
This is evident from Fig~\ref{fig:hst_jwst_comp}.
This indicates that NIR photometry complements HST UV/optical photometry effectively, and that the SED fitting models are able to robustly recover the parameter distributions for the majority of SCCs using HST photometry alone. 
However, for a considerable fraction of SCCs, we still observe significant changes in the derived parameters after NIR photometry is included.
Specifically, $\sim$22$\%$ and $\sim$13$\%$ of SCCs show changes greater than 0.5 dex in age and mass estimates respectively whereas $\sim$40$\%$ SCCs exhibit differences in $A_V$ values exceeding 0.5.
In the following paragraphs, we investigate the SCCs that exhibit such large variations in detail and explore the underlying causes. 
Before that, we first examine the general trends in SCC properties that remain consistent across both the HST-only and HST+JWST runs.

In both runs, the SCCs are clustered in the age bins of young clusters from 5-10 Myr and also the older ages from 100 Myr-1Gyr (top panel of Fig~\ref{fig:hst_jwst_comp}). We notice a considerable reduction in the number of intermediate age clusters from 10-100 Myr.
For dust extinction, $A_V$ values are found to be close to zero for most SCCs with a small but considerable $\sim25\%$ of them showing moderately high extinction values of $A_v > 1$ as well (middle panel of Fig~\ref{fig:hst_jwst_comp}). 
The cluster mass distribution  is skewed towards the low mass end with most of the SCCs falling into the mass range of $10^{3.5}-10^{4.5}$ \msun\ with only a minority present at the higher mass end, i.e. $m > 10^5$ \msun\ (bottom panel of Fig~\ref{fig:hst_jwst_comp}).
A detailed analysis of these distributions and their spatial variations is presented later in Sections \ref{subsec:age_vs_mass} and \ref{subsec:spatial_scc}. However, these trends remain consistent across both {\sc Cigale} runs irrespective of the inclusion of JWST photometry.
An example of a young ($\sim 5$ Myr) and an old ($\sim$ a few 100 Myr) SCC that showed consistent cluster properties across both JWST+HST and HST only {\sc Cigale} runs are shown in the top and middle panels of Figure~\ref{fig:cigale_example}.
In the following paragraphs, we examine how the parameters differ between these two runs and explore the underlying reasons for these variations.

There are several difficulties associated with accurately estimating the star cluster properties using SED fitting methods.
The extent of photometric coverage can significantly affect the best-fit values of important physical parameters such as age ($\tau$), mass ($m$) and dust extinction ($A_{V}$).
As shown in \citet{whitmore2020}, even the choice of the combination of filters used can sometimes  affect the SED fitting considerably. 
Also, it is important to note that our results may be influenced by the choice of input models and parameter ranges used in the SED fitting (see Table~\ref{tab:cigale_tab}). 
For a detailed discussion on how some of these choices affect star cluster SED fitting, see \citet{turner2021}. We also examine the effects of varying a few key parameters, such as metallicity, in Section~\ref{subsec:input_vary}.
Additionally, we address possible contamination from shocked gas to nebular emission lines that can affect SED fitting. The background subtraction with annuli placed close to the source (3–5 pixels, see subsection \ref{subsec:data}), should effectively remove diffuse shock contributions. Moreover, the use of ten filters spanning 3000\AA\ to 3.5$\mu$m further mitigates such effects, as shocks impact only small portions of the SED compared to the broad cluster emission.

However, one of the main challenges remains the age–reddening degeneracy.
This degeneracy refers to the difficulty in distinguishing between a younger SCC with high dust reddening (say, $\tau \leq 10^7$ yr, $A_v \geq 1.0$ ) and an older SCC (say, $\tau \geq 10^8$ yr, $A_v \sim 0$) using photometric methods.
In UV/optical bands, both scenarios can result in similar SEDs with a characteristic dip in the \U\ flux relative to the other long wavelength bands.
In a young, dust-reddened cluster, the dip arises due to strong dust attenuation at shorter wavelengths whereas in an older cluster, the dip is instead caused by
the cluster's integrated light becoming dominated by cooler, lower-mass stars as massive, hot, blue stars evolve off the main sequence.

Conventionally, H$\alpha$ emission serves as a strong indicator of young SCCs with ongoing star formation and helps to break the degeneracy. However, its use in the case of SQ is complicated by two main factors:
(1) As noted earlier, the narrow-band H$\alpha$ images are not sufficiently deep to detect the low-luminosity SCCs found in our catalog. Even with a strong detection, the accurate background estimation becomes challenging due to the presence of widespread diffuse emission across many regions of SQ.
(2) More importantly, large kinematic difference ($\sim \text{a few } 100$ \kms) between the intruder and the rest of the system leads to [N II]$\lambda$6584 from one kinematic component coinciding with $H_{\alpha}$ from the other component in many regions especially along the main shock region \citep{puertas2021}. 
These issues necessitate caution when interpreting narrow-band imaging in SQ. For these reasons, we exclude the H$\alpha$ map from our analysis.
However, we note that the combination of F606W, which includes strong nebular emission lines such as H$\alpha$, and F547M which samples a nearby continuum region free of emission lines, serves as a useful proxy for H$\alpha$ strength. This combination helps greatly in constraining the ages of the SCCs in our SED fitting.

Interestingly, NIR photometry can also play a crucial role in breaking this degeneracy. At longer wavelengths, the two scenarios can produce considerably different SEDs, allowing for a differentiation between young, dust-reddened clusters and older, less obscured ones, albeit with a few limitations.
In the next section, we demonstrate how the inclusion of JWST photometry affects this degeneracy and potentially improves the accuracy of cluster parameter estimation.

\subsubsection{Age extinction degeneracy}

\begin{figure}
    \centering
    \includegraphics[viewport=5 0 1600 1100, width=\textwidth,clip=true]{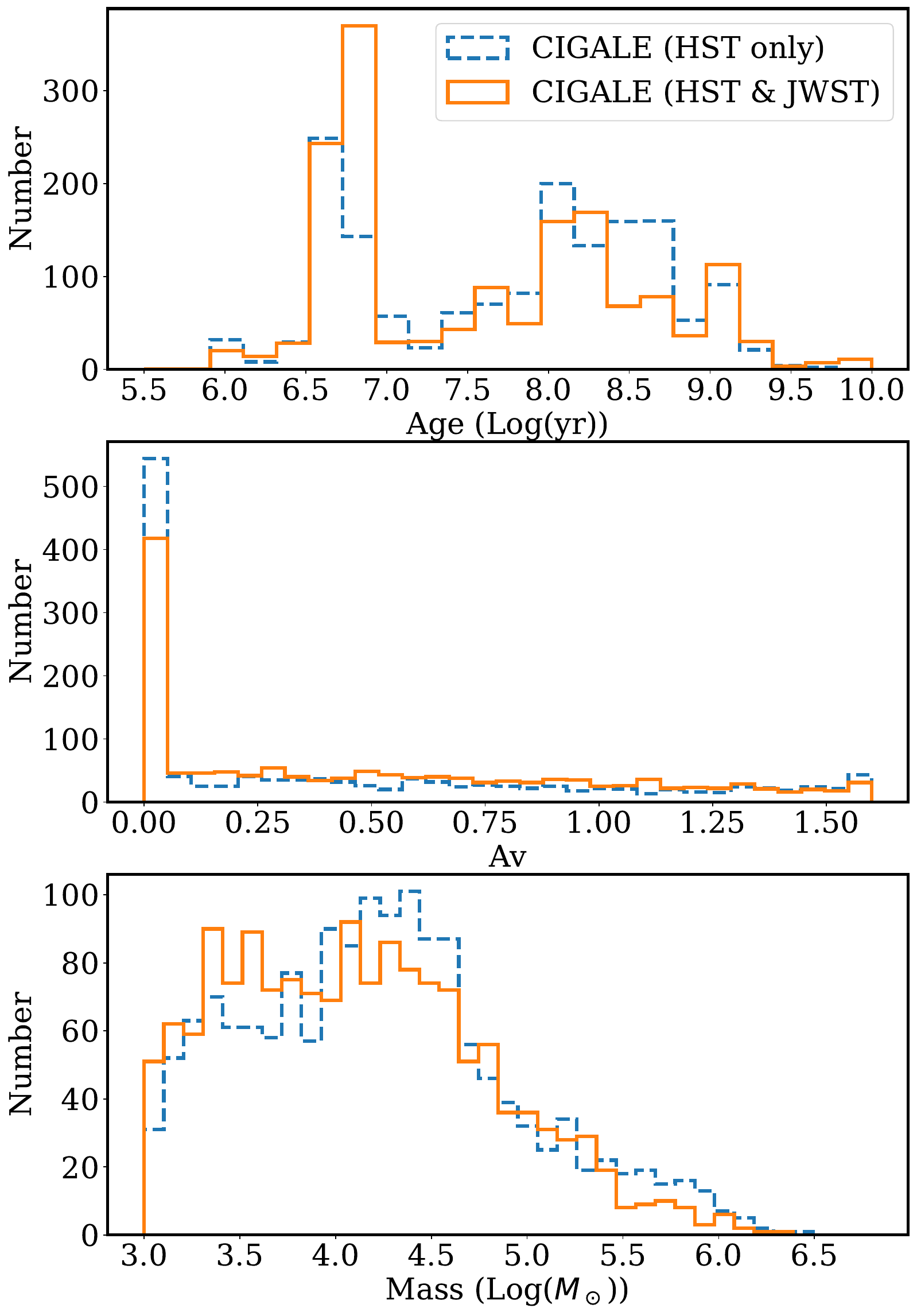}
    \caption{The histogram of the distribution of SCC properties (1,588 SCCs) such as cluster age (top), $A_V$ (middle) and cluster mass (bottom) as derived from {\sc Cigale} fits using only HST photometry (blue dashed lines) and HST+JWST photometry (orange solid line).
    }
    \label{fig:hst_jwst_comp}
\end{figure}

\begin{figure}
    \centering
    \includegraphics[viewport=5 0 2370 2080, width=\textwidth,clip=true]{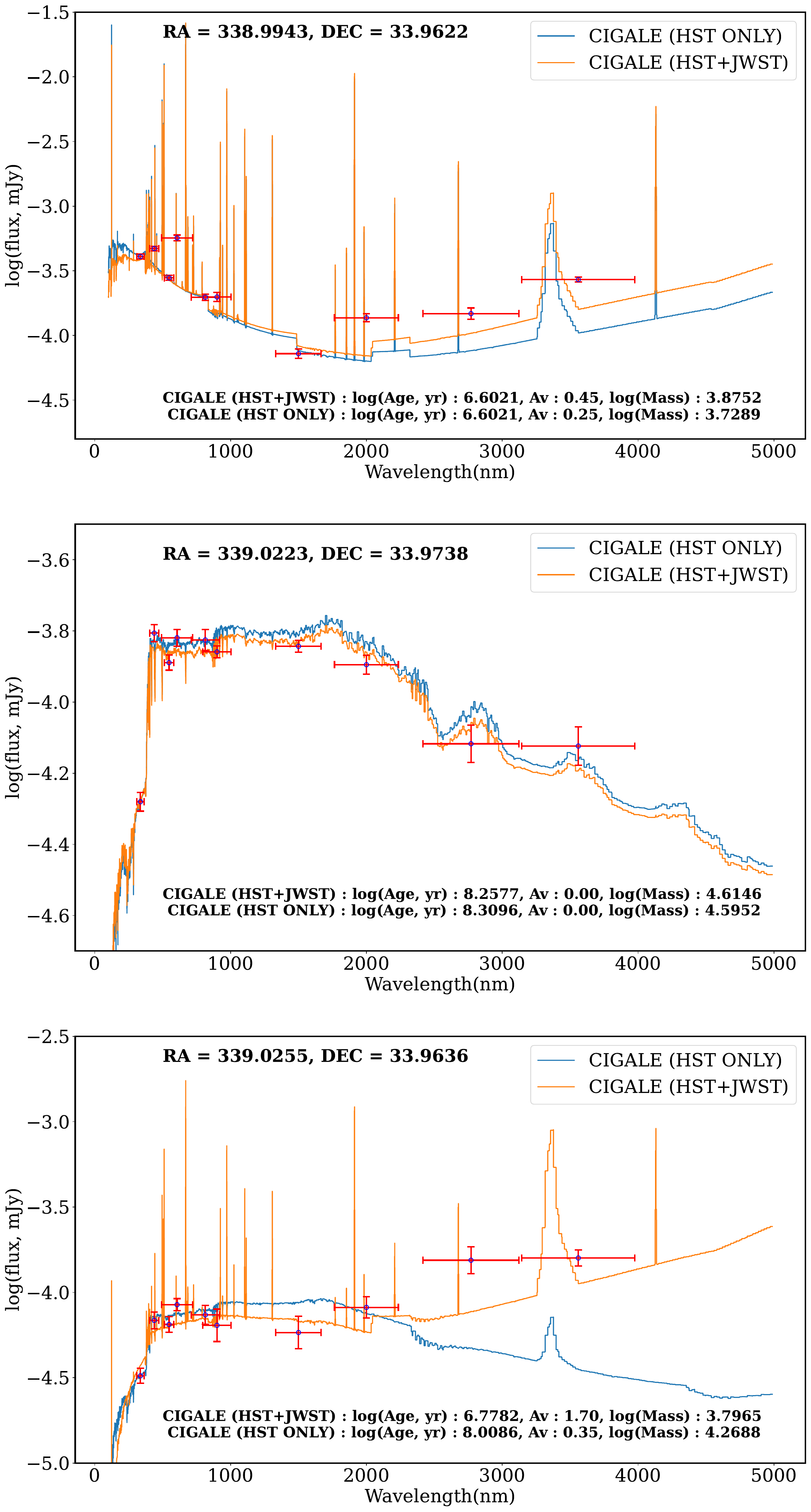}
    \caption{This figure shows three examples of comparison between {\sc Cigale} SED fits and the output parameters based on HST + JWST photometry and HST photometry only for 
    (i) top panel : an SCC with a consistent young age of $\sim 4$ Myr,
    (ii) middle panel : an SCC with consistent older age of $\sim 2 \times 10^8$ yr,
    (iii) bottom panel : an SCC (RA = 339.0255, DEC = 33.9636) which shows large changes in age and extinction ($\Delta \sim$ 1 dex) values after the inclusion of NIR photometry. This is a clear example how NIRCam photometry affects age-extinction degeneracy.}
    \label{fig:cigale_example}
\end{figure}

We examine how NIR photometry affects the derived properties of SCCs by comparing our {\sc Cigale} runs using HST+JWST photometry with respect to HST-only photometry, as described in Section \ref{subsec:cigale}.
As mentioned before, Figure~\ref{fig:hst_jwst_comp} presents histograms of the best-fit values of cluster age ($\tau$), mass ($m$), and dust extinction ($A_{v, ISM}$) for the 1,588 SCCs, both with and without the inclusion of JWST NIRCam photometry.
It is clear that these parameter distributions are considerably affected after incorporating NIR photometry.
Both the ages and masses of several clusters tend to shift toward lower values, while the $A_{V}$ estimates increase, suggesting higher dust extinction in these cases.
Notably, from the top panel of Figure~\ref{fig:hst_jwst_comp}, we find that 121 SCCs classified as older ($\tau \geq 10^{8}$ yr) from HST photometry only are changed to younger ages ($\tau \leq 10^{7}$ yr) when NIR fluxes are included.
This shift is often accompanied by an increase in $A_V$ (middle panel of Figure~\ref{fig:hst_jwst_comp}), highlighting the role of NIR photometry in breaking the age-reddening degeneracy.
We observe that for $\sim40\%$ of the SCCs, the $A_V$ values change significantly (with $\Delta A_V \ge 0.5$) and interestingly, this subset also includes several SCCs where the age estimates remain largely unchanged.
This underscores the importance of NIR photometry in accurately estimating dust properties in star clusters. 
However, this approach is not without limitations, and we discuss several potential caveats in detail in Subsection~\ref{subsec:discuss1_NIR}. 
Now, we explore these trends in more detail in the following paragraphs.

To demonstrate our results, in the bottom panel of Figure~\ref{fig:cigale_example}, we present an example of the SED fit for an SCC in the sample (RA = 339.0255, DEC = 33.9636), where a significant change in the best-fit parameters is observed with the addition of NIRCam filters.
In this case, the estimated age decreases from a few tens of Myr to just a few Myr ($\Delta \tau \sim -1$ dex), while the dust extinction $A_V$ increases by nearly a factor of 7. Furthermore, the cluster mass ($m$) decreases by approximately 0.4 dex.
This motivates us to study the factors that possibly affect the NIR part of the SED and how it affects the estimation of physical parameters as shown in the bottom panel of Fig~\ref{fig:cigale_example}.

Firstly, F150W-F200W color can indicate the presence of young clusters as the F150W traces the starlight continuum devoid of emission lines whereas F200W contains prominent lines such as Pa$\alpha$$\lambda$18750. This is clearly evident in Fig~\ref{fig:cigale_example}.
Next, we examine the NIRCam LW filters such as F277W and F356W bands. For moderate dust extinction values, say $A_V > 1$, we notice a steady increase in the observed NIR flux towards wavelengths longer than $\sim$2$\mu m$ as shown in the bottom panel of Fig~\ref{fig:cigale_example}. This is primarily due to the prominent dust continuum emission in these clusters, which leads to larger flux as we move to longer wavelengths in NIR filters.
{\sc Cigale}, which follows the energy balance principle, accounts for this effect by redistributing the absorbed UV-optical light into the infrared regime \citep{cigale2019}. Hence, the inclusion of F277W and F356W can significantly improve the $A_V$ estimates in our SED fits and further helps in breaking the age-reddening degeneracy in many cases.
Interestingly, F356W coincides with the 3.3 $\mu$m polycyclic aromatic hydrocarbon (PAH) emission feature. This feature is particularly prominent in regions with high dust extinction, which is common for embedded, very young clusters. As a result, young, dust-enshrouded clusters exhibit enhanced flux in the F356W band compared to other NIR bands.
This can be seen in the bottom panel of Fig~\ref{fig:cigale_example} where the simulated PAH emission from {\sc Cigale} clearly leads to higher fluxes around 3.3 $\mu$m.
Note that, in our {\sc Cigale} runs, \citet{dale2014} model is being used to create dust continuum and PAH emission features such as the one at 3.3 $\mu$m.

It has been estimated that the 3.3 $\mu$m PAH emission can cause a difference of up to $\sim$0.12 mag in the measured F200W – F356W colors \citep{linden2023}. In a detailed study of 17 NIRCam-detected young massive clusters (YMCs) in VV114, \citet{linden2024} found that very young and highly dust-enshrouded clusters exhibit both high F150W–F200W and F335M–F200W colors, placing them in a distinct region of the F150W–F200W vs. F335M–F200W color-color diagram (see Figure 3 therein).
However, it is good to keep in mind that F356W flux might be considerably affected by the contribution from the background emission from hot dust (especially in the shock regions) and also from the uncertainties on the strength of 3.3 $\mu$m emission despite the model used in {\sc Cigale}.

We consider potential issues with using photometric flux from long-wavelength (LW) filters such as F277W and F356W in SED fits. These problems arise due to (i) poor spatial resolution of NIRCam LW filters (0.063\arcsec/pixel) and (ii) larger PSFs (FWHM$\sim$0.1\arcsec, see Table~\ref{tab_obs}) at longer wavelengths, which can lead to contamination from both nearby sources and background emission in crowded fields. This can sometimes result in fluxes being overestimated in the F277W and F356W bands.
To check whether the F277W and F356W fluxes are solely responsible for breaking the degeneracy, we performed {\sc Cigale} runs using only the short-wavelength filters i.e. F090W, F150W, and F200W and examined if this still leads to significant changes in the age and $A_V$ parameters.
We found that nearly 60$\%$ (72/121) of SCCs that shifted from older ($\geq 10^{8}$ yr) to younger ($\leq 10^{7}$ yr) ages still exhibited the same age shift, even without the F277W and F356W filters.
This underscores the importance of even short-wavelength NIRCam filters which can greatly affect the age-$A_V$ estimates.

Furthermore, to test the impact of incomplete JWST coverage on SED fitting, we compared 1,077 SCCs detected in all JWST bands with 511 SCCs that have at least one non-detection in JWST bands. Both groups show similar distributions of HST-only SCC properties and a comparable fraction of sources that underwent significant age revisions ($7–9\%$ with $|\Delta \tau| \geq 1$ dex) from HST+JWST runs. However, when three or more JWST bands are undetected (171 SCCs, $\sim10\%$ of the total), the ability of NIR photometry to influence the results, such as breaking the age–extinction degeneracy, is notably diminished and should be taken into account when analyzing this subsample of sources.

With improved estimates on SCC properties, we now examine how these different properties are related to each other and how they vary across SQ in the subsequent subsections.
We use the best-fit parameters obtained using HST+JWST photometry with fiducial parameters (as given in Table~\ref{tab:cigale_tab}) for further analysis in this paper.

\subsection{Varying input parameters}
\label{subsec:input_vary}
In this subsection, we explore the effects of varying certain input parameters in the {\sc Cigale} run that were initially kept constant in the fiducial model, as described in \ref{subsec:cigale}. The goal is to assess how these parameters influence the SED fitting results and to identify any variations observed in the derived properties.

\subsubsection{Extinction curve}
\label{subsubsec:mu}

\begin{figure}
    \centering
    \includegraphics[viewport=5 0 1620 1080, width=\textwidth,clip=true]{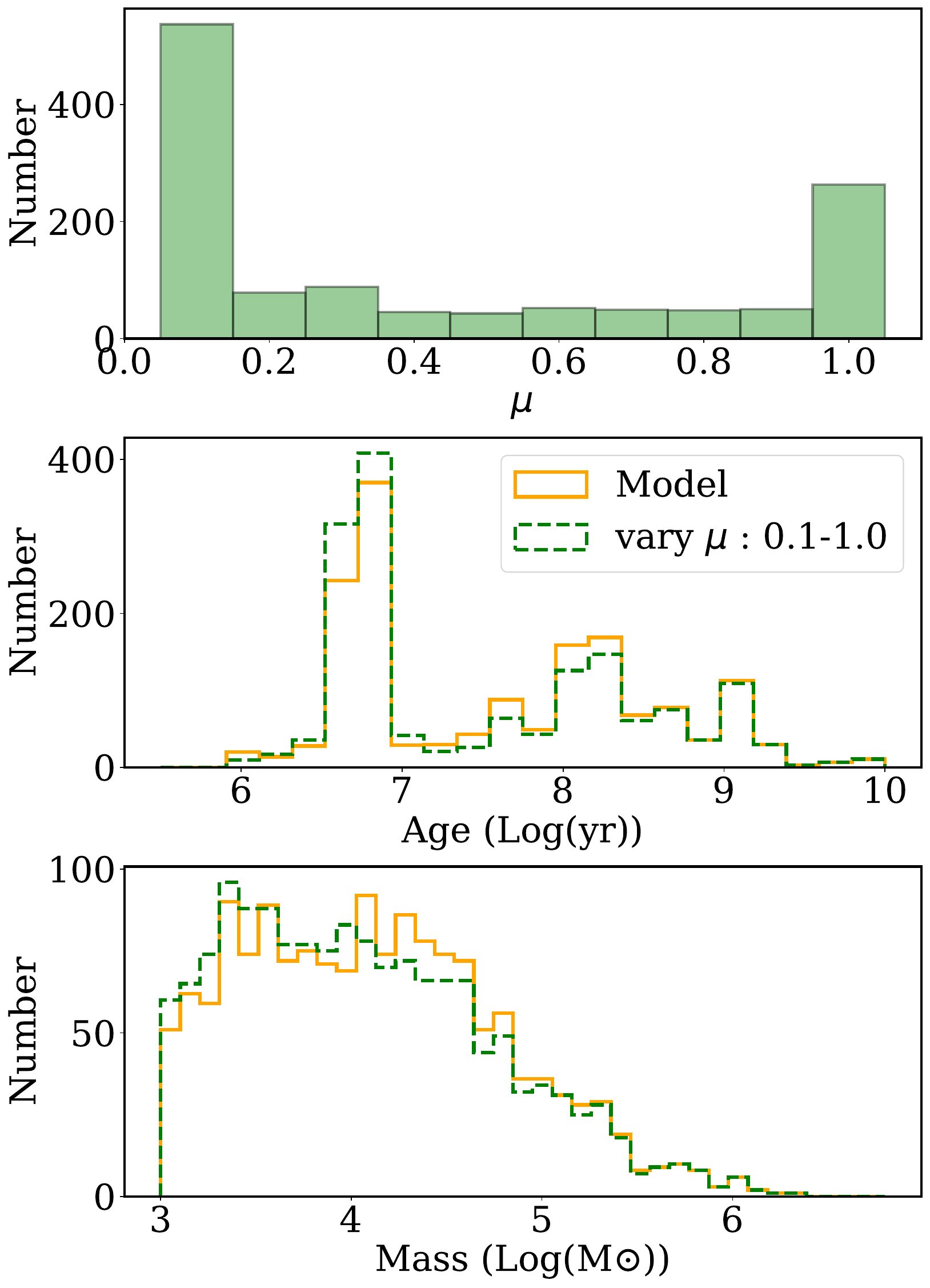}
    \caption{ Top panel : The histogram of best-fit $\mu$ values ($\mu = \frac{A_{V, \mathrm{ISM}}}{A_{V, \mathrm{BC}} + A_{V, \mathrm{ISM}}}$) after keeping it as a free parameter in {\sc Cigale} runs as discussed in \ref{subsubsec:mu}. Note that we exclude sources with $A_{V, total} = A_{V, ISM} + A_{V, BC} = 0$ (334 SCCs) for which $\mu$ values have no physical significance. Middle and Bottom panels : Comparison of {\sc Cigale} parameters such as cluster age (top) and cluster mass (bottom) using fiducial model parameters and after varying input parameters such as $\mu$. 
    }
    \label{fig:cigale_input_vary}
\end{figure}

One particularly interesting input parameter that can be varied in {\sc Cigale} runs is defined as $\mu = \frac{A_{V, \mathrm{ISM}}}{A_{V, \mathrm{BC}} + A_{V, \mathrm{ISM}}}$, where $A_{V, \mathrm{ISM}}$ and $A_{V, \mathrm{BC}}$ represent the V-band attenuation in the interstellar medium and in the birth clouds of star-forming regions, respectively \citep{cigale2019}.
Note that in this case, the total attenuation is given by $A_V = A_{V, \mathrm{BC}} + A_{V, \mathrm{ISM}}$, whereas in the fiducial run, it was simply $A_V = A_{V, \mathrm{ISM}}$.
These attenuations differ as the birth clouds (BC) are assumed to have a much steeper extinction with a power-law slope of --1.3 whereas for ISM, the slope is --0.7.
As mentioned in \ref{subsec:cigale}, the fiducial run ignores the effects of  $A_{V, BC}$ and keeps $\mu =1$ where only ISM contributes to dust attenuation. 
Since radiation from young stars must pass through both the birth clouds and the ISM, {\sc Cigale} allows for the implementation of a two-component attenuation model in which stars younger than 10 Myr are affected by both the birth cloud and ISM extinction curves in these modified runs.
In contrast, stars older than 10 Myr are only affected by the ISM extinction curve.
This means, as $\mu$ approaches zero, the birth clouds contribute more to the extinction and this leads to a steeper extinction curve. 

We now examine how $\mu$ varies and affects other SCC parameters when treated as an additional free parameter. 
To test this, we performed a {\sc Cigale} run varying $\mu$ from 0 to 1 in steps of 0.1, while keeping all other input parameter grids identical to those in the fiducial run. 
Any significant deviation of the best-fit $\mu$ values from $\mu = 1$ would indicate that dust attenuation cannot be fully explained by the ISM alone in these SED fits, suggesting a non-negligible contribution by other means such as birth cloud extinction.
The results from the {\sc Cigale} run, as shown in the top panel of Figure~\ref{fig:cigale_input_vary}, clearly indicates $\mu$ is highly variable and the best fit SEDs mostly prefer $\mu$ values close to zero despite the fiducial value being one.
We note that SCCs with $A_{V,\mathrm{total}} = A_{V,\mathrm{ISM}} + A_{V,\mathrm{BC}} = 0$ (334 SCCs) are excluded from this analysis, since $\mu$ has no physical meaning in these cases.
We find $\sim$16 $\%$ (259/1588) of SCCs are fitted with $\mu =1$ while $\sim$30$\%$ (500/1588) of SCCs give $\mu = 0.1$.
This means that a considerable number of SCCs might need a steeper extinction curve, possibly due to extinction originating from dense molecular clouds around star-forming regions.

Furthermore, we check if the variations in $\mu$ lead to any changes in the best-fit values of cluster age and mass as shown in the middle and bottom panels of Fig~\ref{fig:cigale_input_vary} respectively.
The best fit SCC properties from this {\sc Cigale} run is provided as a separate column in Table~\ref{tab_cigale}.
We find that the overall age and mass distributions remain largely unchanged, except for a subset of $\sim$70 (4$\%$) SCCs where the ages show a large decrease from $>10^8$ yr to $<10^7$ yr, accompanied by a corresponding decrease in $\mu$.
This indicates that the nature of dust properties, specifically the steepness of extinction curve, can affect the cluster properties and should be seriously considered while performing SED fitting routines in future analyses.

A detailed analysis of SCCs with lower $\mu$ values ($\mu <$ 0.5) do not show any correlation with other cluster properties and are distributed across SCCs of all ages and masses. In most cases, these reduced $\mu$ values are favored as they provide better fits by reproducing the large \U-\B\ color driven by the lower \U\ fluxes strongly affected by extinction. Projection effects may also contribute, as some SCCs could be affected by unrelated dust along the line of sight given the chaotic nature of the system.
However, we acknowledge that introducing this as an additional free parameter can increase the overall uncertainties in other key parameters, and therefore it was excluded from our fiducial runs.

\subsubsection{Metallicity}

\begin{figure}
    \centering
    \includegraphics[viewport=5 0 1000 340, width=\textwidth,clip=true]{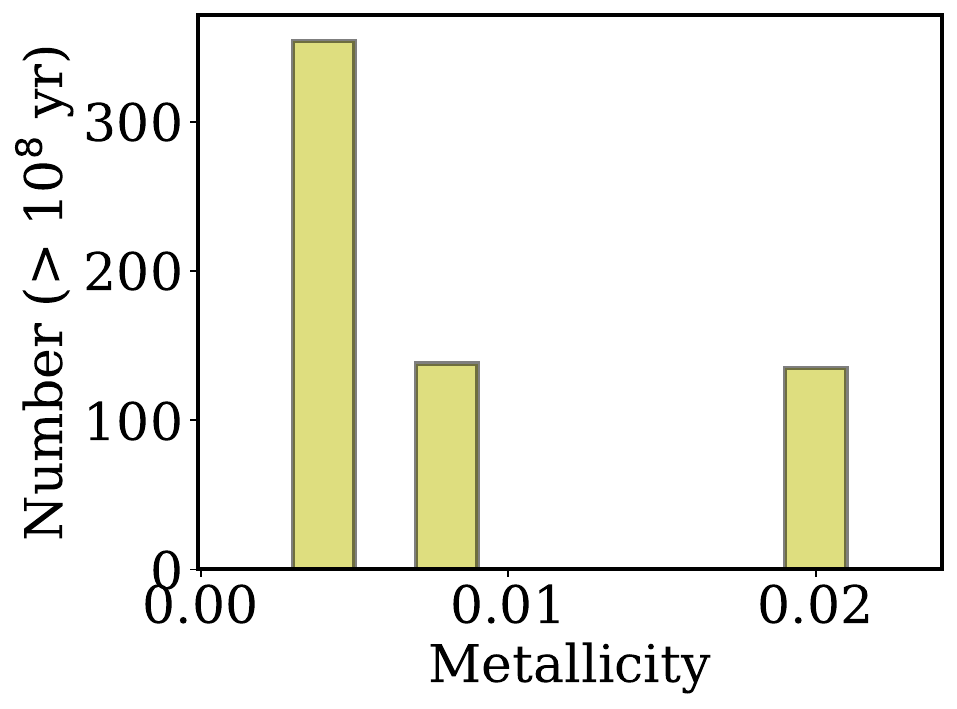}
    \caption{Histogram of best-fit metallicity values after keeping it as a free parameter for the older clusters ($\tau > 10^8 yr$).}
    \label{fig:hist_metallicity}
\end{figure}

Metallicity (Z), defined in mass fraction of the metals in a star as considered in the SSP model, is another important parameter which can affect the SED fit considerably, especially for the older globular clusters.
{\sc Cigale} allows us to run a grid on the metallicity with the possible values of 0.004, 0.008 and 0.02.
The default solar value used in the fiducial run is 0.02.
The older globular clusters are known to be in metal-poor environment due to their early formation and it is interesting to check if they are better fitted with a sub-solar metallicity.
Figure~\ref{fig:hist_metallicity} shows an histogram of best-fit metallicity for older clusters with ages above 10$^8$ yr.
It is clear that older globular clusters prefer sub-solar metallicity as most SCCs have metallicity values less than 0.02, with the peak at 1/5 solar metallicity (Z=0.004).
This is consistent with the studies such as \citet{whitmore2020} that assumed Z=0.004 for the cluster age dating in NGC 4449.
They also demonstrated that varying the assumed metallicity by a factor of a few does not considerably affect the age estimates across the full age range (see Figure 15 therein). 
Consistent with this, changes in metallicity do not produce substantial differences (i.e., $>0.5$ dex) in the age and mass estimates for $\sim93\%$ and $\sim98\%$, respectively, of the older clusters ($\tau > 10^8$ yr) in our sample.

\subsection{Cluster age vs mass relations}
\label{subsec:age_vs_mass}

\begin{figure}
    \centering
    \includegraphics[viewport=0 0 550 260, width=\textwidth,clip=true]{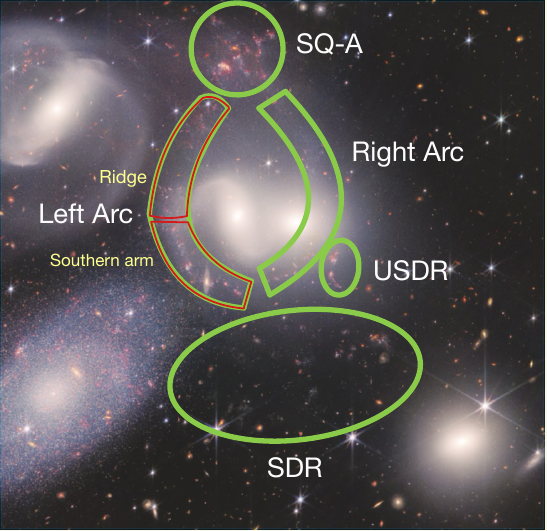}
    \caption{The zoom-in of the JWST NIRCam false color image, focusing on a region of interest that is mostly covered by \U\ and \Vm\ HST bands which has the smallest field of view. The different regions of interest in the SQ field as mentioned in \ref{subsec:age_vs_mass} are marked here for clarity.
    }
    \label{fig:hst_uband_fov}
\end{figure}

\begin{figure}
    \centering
    \includegraphics[viewport=5 0 1550 860, width=\textwidth,clip=true]{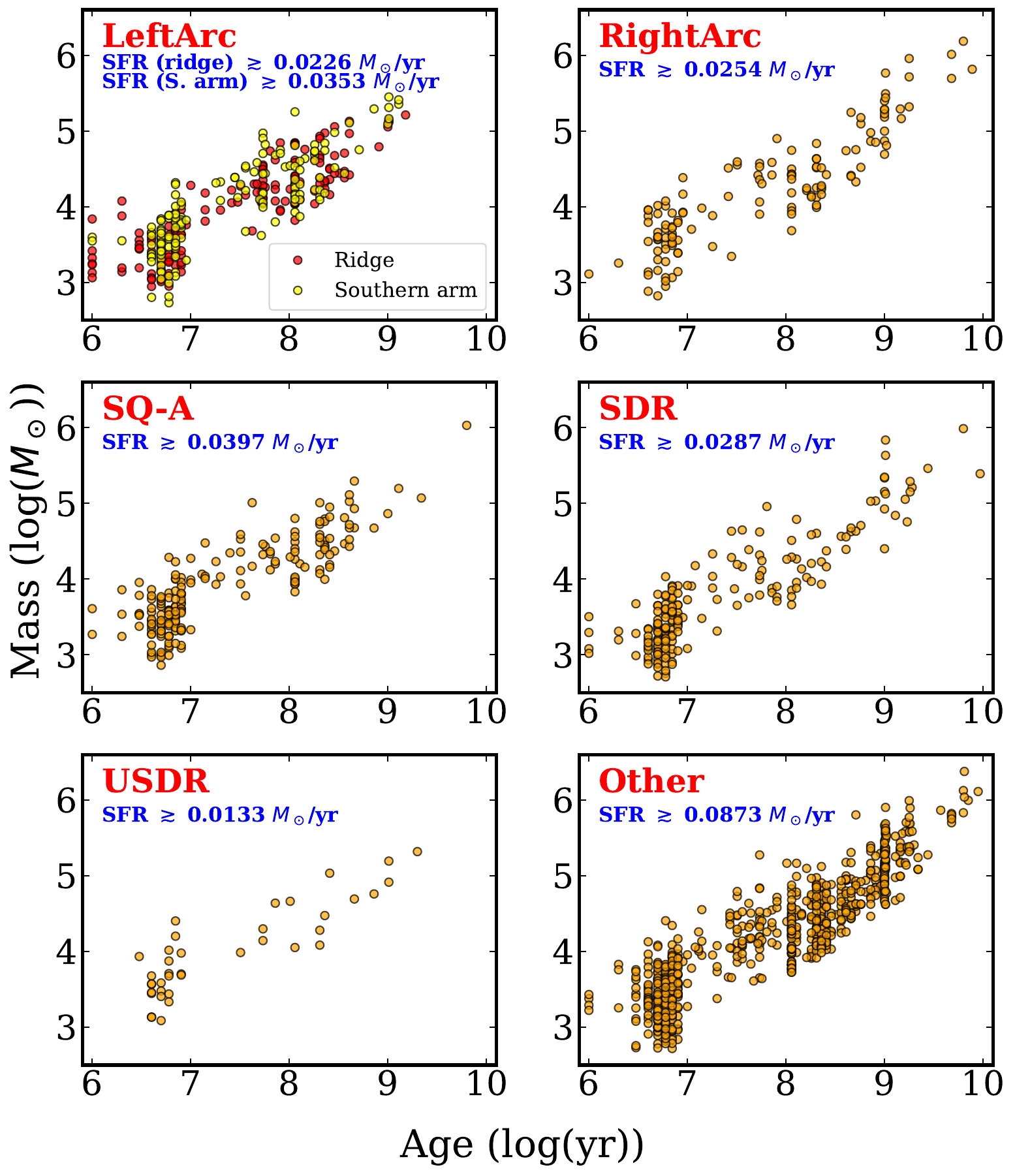}
    \caption{The cluster age vs cluster mass relation as obtained from {\sc Cigale} SED fitting for different regions of interest as described in \ref{subsec:age_vs_mass}. The star formation rate estimated for the regions based on derived age and mass as explained in \ref{subsec:spatial_scc} is shown in each panel.
    }
    \label{fig:age_vs_mass}
\end{figure}

In this subsection, we analyze the relationship between two key SCC properties - cluster age and mass - as derived from the {\sc Cigale} runs using HST+JWST photometry. 
We plot the logarithmic age versus mass diagram to examine overall trends and investigate potential spatial variations in this relation across different regions of the SQ field.
For this, we divide the SQ field into several regions of interest, namely Left arc, Right arc, SQ-A, Southern Debris region (SDR), Upper SDR (USDR) and an ``other" category which include the rest of the field.
These regions are shown in Figure~\ref{fig:hst_uband_fov} for clarity. 
The Left arc denotes the eastern tidal arc close to NGC 7318b, which encompasses the strong shock region \citep{Allen1972, Xu2003} whereas the Right arc represents the Western tidal arm that likely originates from 7318b but overlays 7318a.
The Left arc can be further divided into (i) a "ridge" structure which is the main shock front that crosses SQ and strongly multiphase in nature \citep{guillard2022} and (ii) NGC 7318b's prominent arm to the east (hereafter referred as ``Southern arm" for convenience) where it coincides with the ridge and extends further towards south as shown in Fig~\ref{fig:hst_uband_fov}.
This division is further motivated by the clear kinematical difference between the molecular gas in these two regions \citep{emonts2025}.
The SQ-A is a region towards north of 7318a/b where the two tidal arcs intersect and also show strong H$_{\alpha}$ emission indicating the presence of young star clusters \citep{Xu1999, Gallagher2001, Lisenfeld2002}.
The SDR region, towards the south of 7318a/b,  also hosts a number of young star clusters and has a significant amounts of neutral hydrogen present \citep{Williams2002}.
The upper SDR is a small overdensity of young clusters which is also suspected to be of tidal nature as well.
For more details, we refer the reader to \citet{fedotov2011}, which provides a comprehensive description of these regions in their analysis.

Each subplot in Figure~\ref{fig:age_vs_mass} represents the above-mentioned regions. This will help us to understand if star cluster properties change depending on where they are located in the field.
Fig~\ref{fig:age_vs_mass} reveals several interesting general trends. 
Firstly, a strong correlation between masses and ages (as observed in the figure) is expected since star clusters dim as they age and only the brightest (massive) clusters are detected, particularly at late ages. 
However, what is surprising is the complete absence of massive ($>10^4$ \msun) young clusters ($< 10^7$ yr) (YMCs) in all regions of SQ.
This can be due to several reasons, (1) we may be missing out on young massive clusters still in the embedded stage, hence missed in UV bands in HST, 
(2) Clumpy, porous dust structures around the star clusters will lead to only a certain fraction of the light reaching us, making them appear less massive  and also affecting our estimates on age and dust extinction \citep{kelsey2009},
(3) Another possible reason comes from the work of \citet{Rodruck2023}, who studied the nearby interacting system NGC 1487. This system—likely a merger of dwarf galaxies—shows a similar absence of clusters more massive than $10^4$$M_\odot$, despite having numerous lower-mass ($\sim10^{3.1}$$M_\odot$) clusters. \citet{Rodruck2023} hypothesized that the low ambient pressures in dwarf-dwarf mergers may suppress the formation of massive clusters. 
However, it is interesting that we observe the same effect despite 7318a/b being a massive interacting pair and also note that a complete lack of YMCs is not commonly observed in star cluster studies of other nearby galaxies.
Additionally, we note that a majority of the massive clusters with masses above 10$^5$ \msun\ are older than 10$^8$ yr in all the regions.

Also, there is a relative absence of older clusters (between ages of 10$^9$ and 10$^{10}$ yr) in the Left arc, SQ-A, and USDR regions. The presence of such clusters in the Right arc and SDR can possibly be explained by their proximity to NGC 7318a, which is classified as elliptical \citep{konstant2014} and is expected to have a significant population of older clusters.
Additionally, we observe a recent star formation episode at $\sim$10$^{6.5}$ yr in all five regions, evident from an overdensity of clusters at that age. The regions Left arc and SQ-A had roughly continuous star formation starting from 10$^9$ yr up to the present, with an additional noticeable peak at around $\sim$ a few 10$^8$ yr. This trend may also be seen in the Right arc region, although less pronounced.
In both SDR and Right arc regions, a reduction in star formation is observed between 10$^7$ and 10$^{7.5}$ years. The USDR region has the smallest number of SCCs detected, preventing us from drawing any significant conclusions about this region.
We also note that there is significant star formation following similar trends outside the regions of interest considered here, as shown in the last panel of Fig~\ref{fig:age_vs_mass}. 
This suggests that active star formation is not just confined to a few localized regions but is instead more widespread across the SQ field, indicating a more dispersed and extended star formation activity.

{\it To summarize, given different physical and kinematical conditions in the regions where the young star clusters are detected in the intra-group medium, i.e. SQ-A, the strong shock region in the Left arc and SDR, it is  remarkable how similar their distributions are in terms of ages and masses. 
It would appear that star formation is happening rather uniformly on a large scale in the tidal features around the two interacting galaxies.}

\subsection{Spatial distribution of SCCs}
\label{subsec:spatial_scc}

\begin{figure*}
    \centering
    \includegraphics[viewport=-5 0 670 590, width=\textwidth,clip=true]{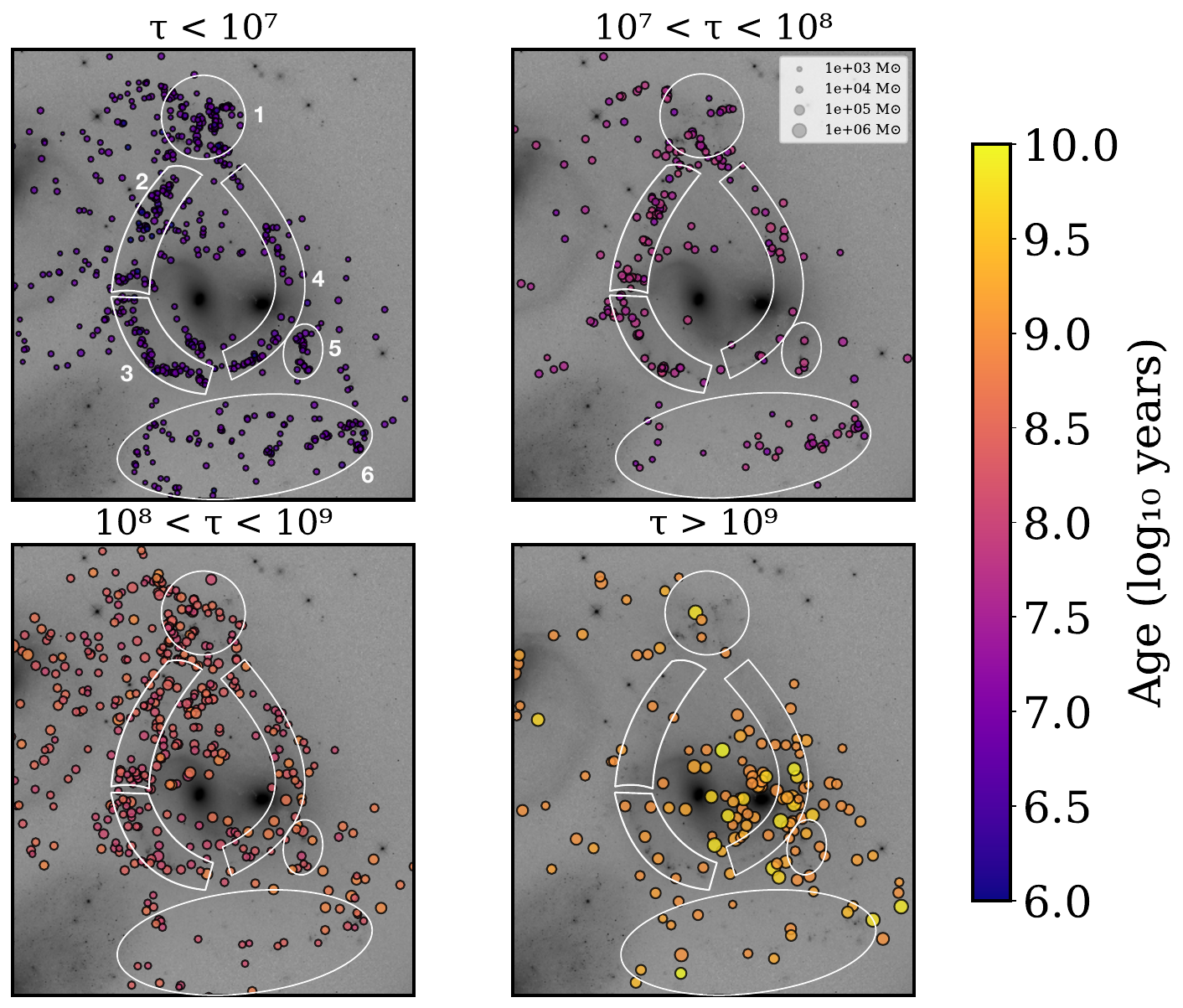}
    \caption{The figure shows the spatial distribution of SCCs around the merging galaxies 7318a/b in different cluster age bins as derived from the {\sc Cigale} SED fits (color coded with respect to the cluster age). The different regions of interest, as shown in Fig.~\ref{fig:hst_uband_fov}, are marked in white: 1- SQA, 2- Ridge (LeftArc), 3- Southern arm (LeftArc), 4- RightArc, 5- USDR and 6- SDR. 
    It can be seen that the very young SCCs ($\sim$a few Myr) are predominantly clustered along shock regions near the merging galaxies, while older globular clusters are more widely scattered.
    }
    \label{fig:age_sq}
\end{figure*}

\begin{figure*}
    \centering
    \includegraphics[viewport=-5 0 660 590, width=\textwidth,clip=true]{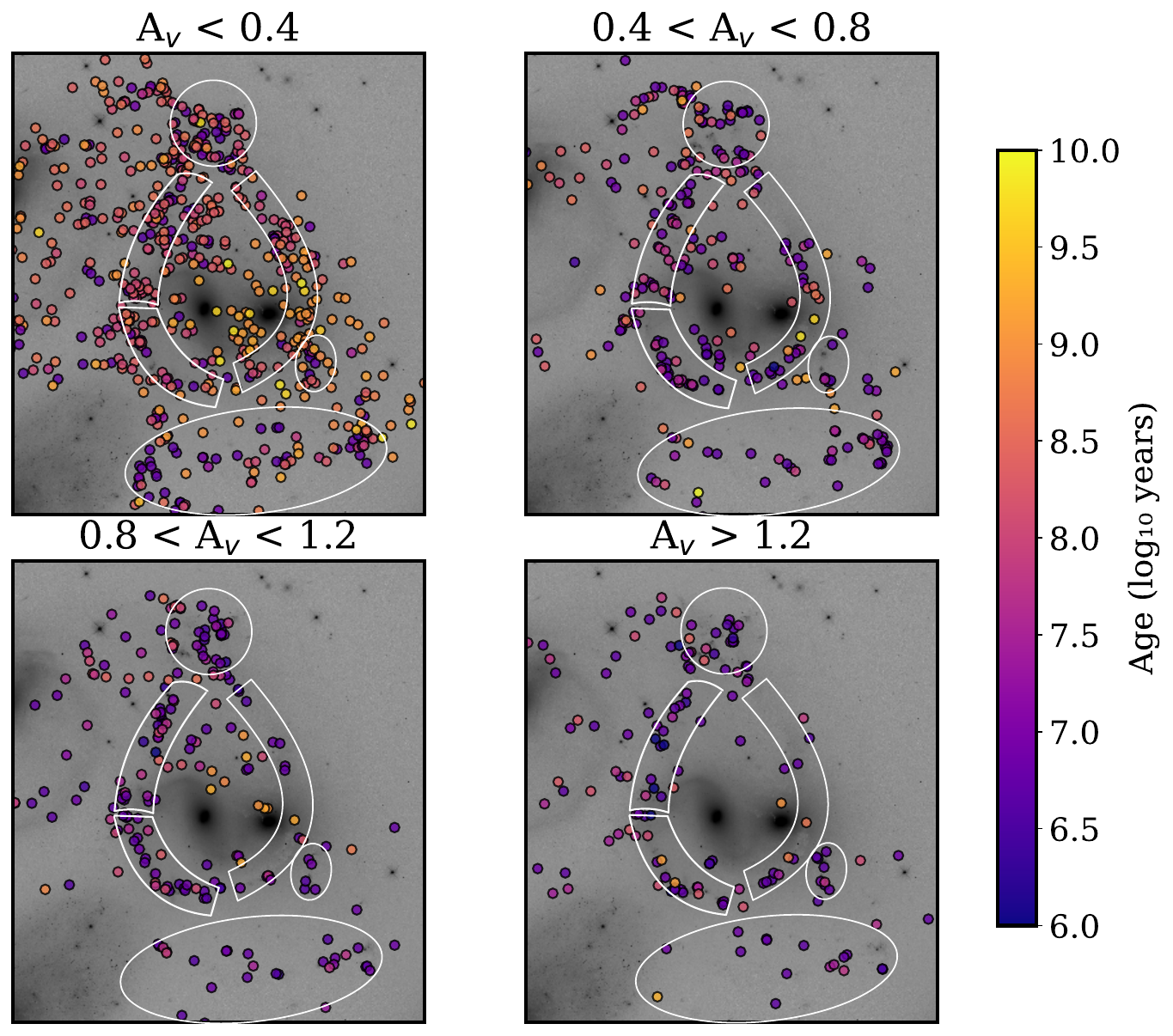}
    \caption{Spatial distribution of SCCs in different $A_V$ bins (color coded with cluster age). The different regions of interest are shown here in white, similar to Fig~\ref{fig:age_sq}.
    }
    \label{fig:Av_sq}
\end{figure*}

In this subsection, we look at the spatial distribution of the SCCs based on their age and dust extinction.
For this, we bin all SCCs in the catalog into four age bins according to their derived {\sc cigale} ages 
($\tau$ $<$ 10$^7$ , 10$^7$ $<$ $\tau$ $<$ 10$^8$, 10$^8$ $<$ $\tau$ $<$ 10$^9$ , and $\tau$ $>$ 10$^9$ ; all ages in years) 
and show them superimposed on the HST \V\ image of the Stephan's Quintet as shown in Figure~\ref{fig:age_sq}. 
In a similar way, we also bin all SCCs according to their estimated dust extinction levels ($A_V$ $<$ 0.4 , 0.4 $<$ $A_V$ $<$ 0.8, 0.8 $<$ $A_V$ $<$ 1.2, and $A_V$ $>$ 1.2)  as shown in Figure~\ref{fig:Av_sq}.

It is evident that the young star clusters ($\tau < 10^7$ yr) are tightly concentrated along the tidal features surrounding NGC 7318a/b and typically have relatively low masses ($M < 10^4$$M_\odot$). The SCCs with intermediate ages ($10^7 < \tau < 10^8$ yr) also show a spatial distribution aligned with the tidal structures, particularly within the Left arc. However, it is noteworthy that the number of such intermediate-age clusters is significantly reduced in regions like the Right arc and SDR.
They are also a bit more massive than the ones in the previous bin, with masses reaching M$\sim$ 10$^5$ \solarM\ and an average mass of M $\sim$ 10$^{4.3}$ \solarM. 
The fact that they still follow the tidal features could mean that the interaction-triggered star formation in the arcs has been going on for a long time ($\sim$ a few tens Myr).
However, the SCCs with ages between 10$^8$ $<$ $\tau$ $<$ 10$^9$ yrs are more scattered and distributed over the entire field. 
Their masses range between 10$^4$ \solarM\ and 10$^{5.5}$ \solarM, with few SCCs outside of that range. 
We find that more massive SCCs are generally located within or near the galaxies. Their relatively uniform spatial distribution around the galaxies suggests that they were likely deposited there, rather than having formed in situ. 
The SCCs in the last bin (with ages over 10$^9$ yr) most likely represent globular clusters predominantly concentrated around the elliptical NGC 7318a. 
The estimated extent for the globular cluster system of NGC 7318a has a diameter of $\sim$ 30 kpc, encompassing NGC 7318b, the majority of the Left arc, the entirety of the Right arc and Upper SDR, and most of the SDR. These clusters represent the most massive population, with masses ranging from 10$^5$ \solarM\ to 10$^6$ \solarM.

Dust extinction as quantified by $A_V$ shows a spatial distribution that is roughly anti-correlated with age as shown in Fig~\ref{fig:Av_sq}.
From the figure, it is clear that the very young SCCs ($\tau$ $<$ 10$^7$ yr) have a moderately higher dust extinction ($A_v > 0.8$) as evident from their spatial distributions in the last panel of Fig.~\ref{fig:Av_sq}.
This suggests that young clusters may have a relatively higher amount of dust present in their immediate environment (possibly from their birth clouds), yet to be cleared out by the stellar feedback effects. 
There are many recent studies showing that dust clearing in young, massive clusters (YMCs) may occur on time scales less than a few Myr \citep{messa2021, linden2023, mcquaid2024}.
The relatively smaller values of $A_V$ (compared to $A_V \sim 5-15$ in the case of YMCs), coupled with the nearly complete absence of very young clusters ($\tau < 3$ Myr) in our sample indicate that the HST-selected SCCs are missing YMCs that are likely only detectable in the NIR \citep{whitmore2023, rodriguez2023, linden2024}.
This again may also explain the lack of YMCs in the age vs mass plots as discussed in \ref{subsec:age_vs_mass}.


The star cluster ages and masses derived from the {\sc Cigale} runs allow us to estimate the star formation rates (SFR) across different regions. We sum the masses of all young clusters with ages $\tau < 10^7$ yr within each region and divide this by  $10^7$ yr to approximate the SFR. This method yields SFRs of 0.0226 \msun/yr, 0.0353 \msun/yr, 0.0254 \msun/yr, 0.0397 \msun/yr and 0.0287 \msun/yr for the ridge (Left arc), Southern arm (Left arc), Right arc, SQ-A, and SDR regions, respectively.
It is important to note that these values likely represent lower limits, as our point source detection algorithm may underperform in regions with substantial diffuse emission, such as SQ-A. Expanding the current catalog to include NIRCam-detected point sources is expected to refine these estimates and also help in detecting YMCs if present in the SQ (Aromal et al., in preparation).

\subsection{Comparison with cold molecular gas distribution}
\label{subsec:molecular_map}

\begin{figure*}
    \centering
    \includegraphics[viewport=-5 0 900 630, width=\textwidth,clip=true]{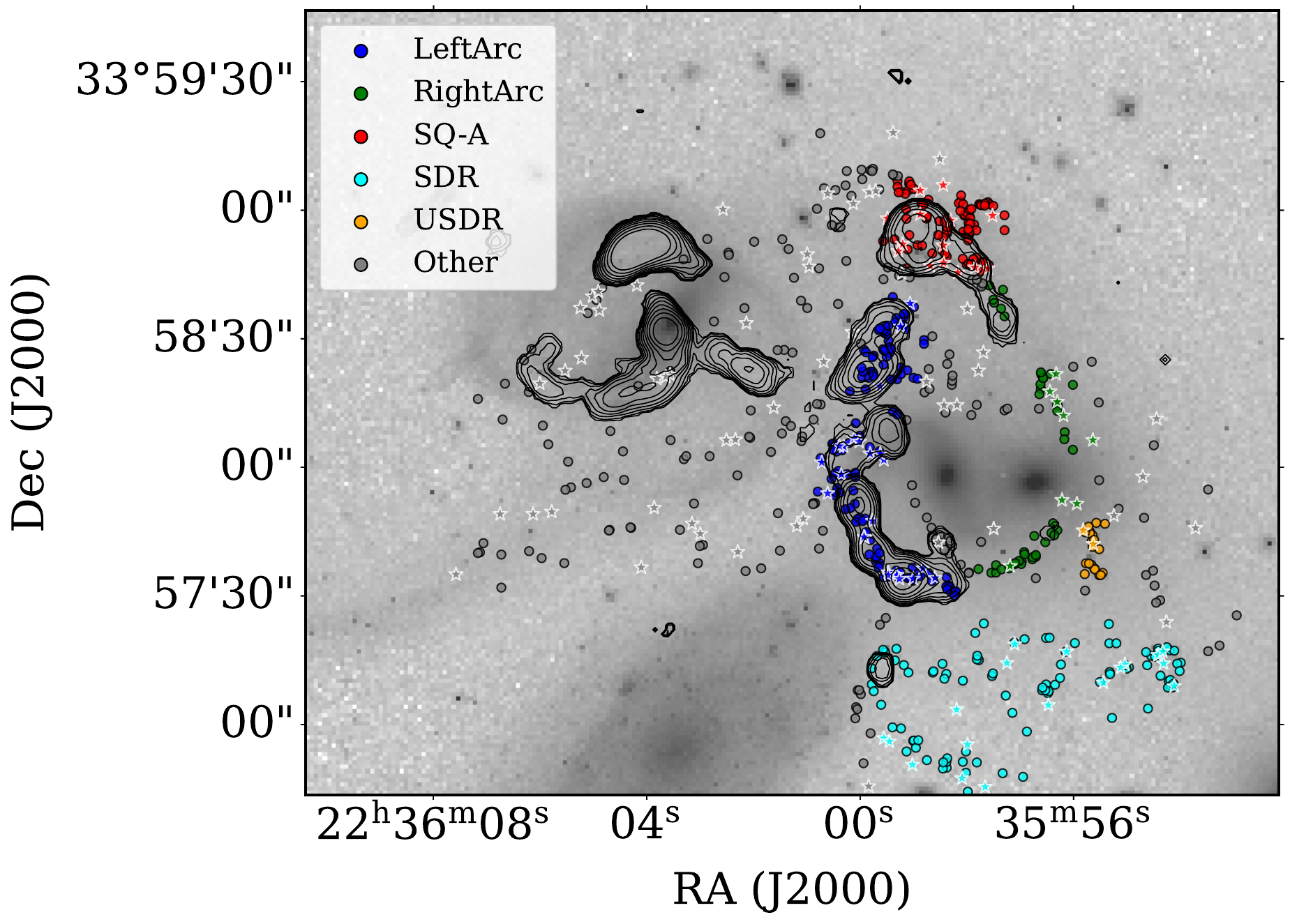}
    \caption{The total intensity (moment-0) map of CO(2–1) molecular emission obtained with ACA \citep{emonts2025}, overlaid on the spatial distribution of young SCCs ($\tau < 10^7$ yr) across the field, color-coded by region of interest. The SCCs whose ages shifted to $<10^7$ yr after including NIR photometry (from initial HST-based estimates of $>10^8$ yr) are highlighted with star symbols, outlined in white for clarity. The contour levels start at 0.4 Jy beam$^{-1} \times$  \kms\ and increase by a factor $\sqrt{2}$. The background is the HST WFC3 image taken in \V\ filter \citep{fedotov2011}. }
    \label{fig:molecular_map}
\end{figure*}

The presence of numerous young star clusters in Stephan’s Quintet suggests a significant reservoir of cold molecular gas within the system. Understanding the efficiency at which this cold gas is being converted into stars is particularly important, especially given the extragalactic nature of these clusters.
Using the Atacama Compact Array (ACA), \citet{emonts2025} recently mapped the large-scale distribution and kinematics of cold molecular gas across the central $\sim70\times70\text{ kpc}^2$ of Stephan’s Quintet through CO(2-1) emission. 
Figure~\ref{fig:molecular_map} shows the contours of the integrated CO(2–1) emission obtained with the ACA resolution of 8\arcsec × 7\arcsec\ ($\sim3.7\times3.2\text{ kpc}^2$), superimposed on the HST \V\ image.
The largest angular scales on which these ACA observations can detect extended emission is $\sim$30\arcsec\ that corresponds to $\sim$14 kpc.
The points on the image represent the young clusters ($\tau < 10^7$ yr) color coded for different regions.
Interestingly, CO emission exhibits different kinematics depending on its location, with velocities ranging between the luminosity-weighted barycentric group velocity of $\sim$6600 \kms\ and the blue-shifted velocities of $\sim$5600 \kms\ associated with the approaching galaxy NGC 7318b (see CO moment-1 map in Figure 2 of \citet{emonts2025}).

From Fig~\ref{fig:molecular_map}, we observe that CO(2-1) emission is mostly concentrated on SQ-A, Left arc and NGC7319 galaxy.
Interestingly, the central regions of NGC7318a/b galaxies do not show any detectable emission.
We find that the young clusters in our sample closely trace the distribution of cold molecular gas, particularly in the Left arc and SQ-A regions. 
This alignment also includes SCCs that were reassigned to younger ages following the inclusion of JWST NIR photometry, as indicated by the star symbols (outlined in white) in Fig.~\ref{fig:molecular_map}, further supporting their likely young nature.

We note that the southern arm region of the Left arc (the prominent arm of 7318b to the east as described in \ref{subsec:age_vs_mass}) is where
the young clusters align closely with the core of the CO emission contours.
This is the only region where the CO(2-1) emission is at a velocity $\sim 5600$ \kms of the intruder galaxy. 
Moreover, this emission also has a red wing that extends to the velocities of the molecular gas on the ridge ($\sim$ 6000 \kms) located north of the Left arc.
Interestingly, we observe that in several regions, such as (i) the ‘ridge’ structure  and (ii) the SQ-A region, young clusters tend to form along the edges of the CO emission contours, in addition to the central regions of peak emission.
In these regions, we observe active star formation 
where the CO(2-1) moment 1 map shows velocities that are between the luminosity-weighted barycentric group velocity of 6600 \kms\ and the blue-shifted velocities of $\sim$5600 \kms\ associated with the approaching galaxy 7318b.
This may indicate the presence of decelerated, shocked gas from the intruder galaxy, which undergoes relatively shorter cooling timescales before forming stars in these regions \citep{guillard2012, emonts2025}.
Interestingly, the central part of the SQ-A shows different CO kinematics compared to other regions in SQ and shows higher velocities close to $\sim$6800 \kms (see Figure 2 in \citet{emonts2025}).
However, it is difficult to associate our SCCs with different kinematic components in the molecular gas because of their poor resolution.
We also check the distribution of older clusters to find that the majority reside outside these regions of strong molecular gas emission.

We observe significant CO emission from NGC 7319 and the ``bridge" which is a 10 kpc long structure 
that extends from the center of the galaxy towards the Left arc.
However, it is clear that this region lacks any young clusters as evident from Fig~\ref{fig:molecular_map} despite the presence of large amounts of molecular gas.
We also note the presence of a bright CO tail to the north of the center of NGC 7319, which contains the strongest CO(2–1) emission in the field. However, we could not include any possible SCCs from this region in our analysis, as it lies outside the area covered in this study.
Interestingly, in the Right arc, SDR, and USDR regions, we found numerous young clusters despite the apparent lack of detectable CO emission (except at the northernmost tip of the Right arc, near SQ-A, where CO emission is detected).
It should be noted that these regions also show relatively smaller $A_V$ ($A_V$ $<$ 0.8) compared to regions such as the Left arc where significant CO emission is detected.
The main bodies of the two merging galaxies 7318a/b also lack any detectable cold gas, which suggests that much of the gas has been stripped into the intergalactic medium as a result of the galaxy collision.
This may have significant implications for star formation in these galaxies, many of which exhibit low current SFRs, a common phenomenon observed in compact group galaxies \citep{ute2017}.
Some of the implications of these observations are further discussed in subsection \ref{subsec:discuss2_SFR}.

\section{Discussion}
\label{sec:discussion}

\subsection{How NIR photometry affects star cluster SED fitting?}
\label{subsec:discuss1_NIR}

It is important to realize how well JWST NIRCam photometry complements that of HST, leading to consistent SCC properties (within 0.5 dex) derived for $\sim$80$\%$ of SCCs with HST+JWST and HST photometry only.
However, as shown in subsection \ref{subsec:age_Av_degeneracy}, the inclusion of NIR filters can also result in significant changes in age ($\Delta \tau > 1$ dex) and A$_V$ estimates for $\sim$8$\%$ of SCCs.
Our study demonstrates how the NIR band colors complement those of crucial HST colors, such as \U-\B, to put tight constraints on dust estimates. 
The fact that our fits are consistent in both regimes -- UV/optical which is affected by dust extinction and the NIR by the dust continuum -- improves the accuracy of the derived SCC properties.
We were also able to show that the steepness of the extinction curve--quantified by an additional free parameter $\mu$ as defined in Section~\ref{subsubsec:mu}--can affect the SED fits, and sometimes the resulting age estimates as well (see Fig.~\ref{fig:cigale_input_vary}).
Now, the majority of SCCs whose ages shifted to $<10^7$ yr after the inclusion of NIR photometry (from initial HST-based estimates of $>10^8$ yr) show several additional indicators of their likely young nature. These include: (1) their location in regions rich in cold molecular gas, as evidenced by strong CO(2–1) emission; (2) newly derived $A_V$ values that are consistent with the average extinction values reported in each region by previous spectroscopic studies \citep{konstant2014, puertas2021}; and (3) their spatial proximity to other young SCCs identified using HST photometry alone. 

However, the inclusion of NIR photometry also introduces several limitations, which we discuss in detail below. It is crucial to understand these effects, particularly in the context of SQ's unique environment, to improve future star cluster studies.
The first limitation concerns the energy balance principle implemented in {\sc Cigale}, which ensures energy conservation by equating the energy absorbed by dust in the UV/optical regime to the energy re-emitted in the infrared. 
Although this assumption generally holds for many systems, it can become problematic if dust is rapidly cleared from the star cluster environment due to stellar feedback effects. 
Recent studies suggest that dust can be efficiently removed within $\sim$2-5 Myr \citep{linden2024, whitmore2025}, suggesting that the energy balance assumption could lead to artificial overestimates of dust content for relatively older clusters.
However, it is important to note that we imposed an upper limit on $A_V$ ($A_V < 1.6$) in our {\sc Cigale} runs to prevent abnormally large extinction estimates.
Also, we keep in mind the special nature of SQ, where substantial star formation is occurring across extended, shock-driven structures in the intergalactic medium. Additionally, projection effects complicate the interpretation, as the light from a given SCC may pass through unrelated dust (even shock-heated dust) along the line of sight, originating from other systems or foreground structures. 
These factors can sometimes introduce uncertainties in the derived dust properties and should be carefully considered in the analysis.

Secondly, stochastic effects can contribute significantly to the IR part of the SED. For example, even a single red super giant (RSG) in the cluster can dominate the flux above 1 $\mu m$ after 5 Myr \citep{gazak2013}.
The same effects can also originate from thermally pulsating AGB stars.
This may be misinterpreted as dust emission, which can sometimes lead to overestimation of A$_V$ resulting in younger ages.
One way to resolve this degeneracy is by looking at the H$_{\alpha}$ emission, but for reasons mentioned in \ref{subsec:age_Av_degeneracy}, this is not a very reliable method given the special nature of SQ and the data in our hand.
Other ways include differentiating between the IR SEDs from possible RSGs and AGBs to that of a dust continuum. This is beyond the scope of the current paper and will be discussed in a future paper.
Additionally, for cooler stars, systematic discrepancies have been found in stellar spectral models such as PHOENIX, which often overestimate the NIR flux when matched to optical data, reflecting limitations in current atmospheric modeling for late-type stars \citep{lancon2021}.
Thirdly, as mentioned before, the clumpy and porous dust structures around star clusters can also affect SCC estimates by underestimating the dust extinction and also making them appear less massive \citep{kelsey2009}.
Now, it is a relevant question to ask  what happened to the young massive clusters (YMCs) in Stephan’s Quintet? \citet{linden2023} used the F150W, F200W, and F356W NIRCam filters to identify YMCs in VV114, a system located at a similar distance to SQ (D $\sim$ 84.4 Mpc). 
Their study demonstrated that high values of both F150W–F200W and F200W–F356W colors are strong indicators of very young, highly dust-enshrouded clusters. Moreover, \citet{linden2021, linden2023} found that 20–60$\%$ of the NIR-detected SCCs in VV114 are missed in UV-optical images, depending on the HST filter used. 
This highlights the critical role of incorporating NIR photometry to achieve a more complete census of young star clusters.
To further explore this connection in the context of SQ, we are in the process of compiling a catalog of NIRCam-detected clusters (Aromal et al., in preparation). 
Preliminary analysis has already identified a few highly reddened sources in SQA that are bright in the NIR but completely absent in the HST images. 
A catalog of such clusters will allow us to further improve the derived age-mass relations and star formation rates discussed in \ref{subsec:age_vs_mass}, while also helping to mitigate some of the current uncertainties.

The fact that most of our HST-selected young SCCs are well fitted with relatively low to moderate extinction values  ($A_v < 1.6$), even after incorporating the NIR part of the SED, suggests efficient and early gas and dust removal driven by feedback processes \citep{reines2008, allison2018, linden2024}. 
As noted earlier, the young clusters in our sample predominantly have ages greater than 5 Myr, suggesting the highly dust-embedded phase likely concludes within this timescale, allowing the clusters to become visible across the UV to optical bands. This is consistent with recent observational findings regarding the early emergence of young star clusters from their birth clouds.
However, caution is needed when interpreting these results. Most studies on cluster emergence timescales have focused on clusters in typical galactic environments, whereas the majority of the clusters in SQ likely form due to large-scale shocks in the intra-group medium. It remains uncertain whether clusters formed in such extreme conditions follow the same evolutionary pathways as those in typical galaxy environments.
This distinction makes SQ a particularly compelling system, offering valuable insights into in-situ cluster formation in the IGM. Studying how interactions between colliding galaxies trigger star cluster formation in these environments is crucial for understanding broader galaxy evolution processes.

\subsection{Formation of star clusters and the history of galaxy interactions in SQ}
\label{subsec:discuss2_SFR}

Another important implication of obtaining improved age estimates for SCCs in the shock regions is the ability to constrain the timing of the shocks and the various interactions between the SQ galaxies themselves. As discussed in Section~\ref{subsec:age_vs_mass}, we find that most young clusters in the Left arc, Right arc, and SQ-A regions formed around $\sim$5 Myr ago (see Fig~\ref{fig:age_vs_mass} and Fig.~\ref{fig:age_sq}). 
If we assume that these shocks triggered star formation--particularly in the Left arc--or that these events are interconnected and occurred simultaneously, this suggests that the primary shock event induced by the intruding galaxy 7318b
likely occurred at approximately the same time, around 5 Myr ago. 
This finding is consistent with earlier multiwavelength studies that analyzed past interactions in Stephan’s Quintet, which indicated that the collision with NGC 7318b likely took place a few Myr ago \citep{Moles1997}.
Interestingly, our age distribution also reveals a secondary, weaker, and more broadly distributed peak at $\sim 2 \times 10^8$ yr (see Fig.~\ref{fig:hst_jwst_comp}), corresponding to a population of relatively older clusters. This timescale coincides with the estimated epoch of the most recent encounter between NGC 7320C and NGC 7319 \citep{Moles1997, Renaud2010, hwang2012}. 
That event is thought to have stripped much of the gas from NGC 7319 and produced the prominent southern tidal tail, which continues to host active star formation. 
It is also possible that this encounter led to tidal interactions with NGC 7318a (prior to the more recent collision with NGC 7318b) which may account for the presence of older clusters observed in the vicinity of that region.
This might also explain the relative absence of intermediate-age clusters ($10^7 < \tau < 10^8$ yr), as this timescale may represent a relatively quiescent period between the major dynamical interactions discussed above.
{\it Taken together, our SCC age estimates appear to align remarkably well with the known timescales of past dynamical interactions in SQ, underscoring the strong connection between star cluster formation and galaxy interactions in this system.}

In the Left arc, \citet{konstant2014} concluded that the actively star-forming regions projected along the main shock (ridge structure) are kinematically associated with the intruder galaxy and the star formation is severely inhibited in the shocked gas in the IGM due to its turbulent and warm gas.
This was based on the finding that much of the \hii\ regions are associated with the intruder velocity whereas a mix of shocked filaments and \hii\ regions are seen in the IGM velocity.
A more recent study \citep{puertas2019, puertas2021} based on SITELLE spectroscopy which is an imaging Fourier transform spectrometer with large FoV ($11\arcmin\times11\arcmin$, 0.32$\arcsec$ per pixel) has also confirmed these findings.
Using SITELLE data, the authors found a total star formation rate of $\sim3.1$ \msun/yr in SQ, primarily concentrated in SQA and SQB (SQB is outside the region considered in this paper), with most regions showing low to moderate extinction and subsolar metallicities. 
The IGM-velocity, shock-affected regions exhibit lower O/H and N/O ratios, suggesting chemically younger, stripped gas, while star-forming clumps like SQA show more evolved enrichment.
Both studies showed that star formation occurs predominantly at the intruder velocities, and although our analysis lacks direct velocity information, it is reasonable to assume that most of the SCCs in our sample are also associated with the intruder velocity component. 
A high-resolution IFU study in near-infrared, targeting these regions would provide crucial insight into the kinematics and help clarify these assumptions.

Additionally, using JWST MIRI and ALMA observations, \citet{appleton2023} showed that certain regions in SQ, especially in the main shock filament (ridge) in Left Arc, have high warm-to-cold molecular gas fraction indicating cold gas being disrupted and the resulting warm gas showing luminous H$_2$ cooling throughout the shock.
This probably explain the relatively lower rate of star formation (see Fig~\ref{fig:age_vs_mass}) in the ``ridge", which is mostly at IGM velocity \citep{emonts2025}.
Interestingly, it can also be noticed that the SCC distribution in the ridge is non-uniformly distributed towards the west side of the CO emission contours, further away from the regions of peak emission  (see Fig~\ref{fig:molecular_map}).
Furthermore, in the southern arm region of the Left arc, \citet{guillard2012} and \citet{emonts2025} noted that most of the molecular gas is kinematically associated with the intruder galaxy. This is where we see the SCCs closely trace peak CO emission (and hence the molecular gas distribution) compared to the ``ridge" structure and this region shows a higher SFR as well.

In the Right arc, as shown in Fig~\ref{fig:molecular_map}, the lack of CO detection despite active star formation may indicate shorter molecular gas dispersal timescales compared to the Left arc. Studies such as \citet{chevance2020, chevance2023} have shown that GMC lifetimes are highly environment-dependent and can vary significantly both between and within systems. The coexistence of molecular gas and star formation typically lasts only a few Myr before dispersed and destroyed by stellar feedback. Although most studies focus on galactic environments, it is reasonable to assume that GMCs in different regions of SQ, such as the tidal arcs and SQ-A, evolve under distinct physical conditions  \citep[see][for a detailed discussion]{Xu2025}. This could potentially lead to shorter GMC lifecycles in the Right Arc and SDR, explaining their relative lack of CO. 
An alternative explanation comes from the presence of very old clusters ($>10^9$ yr; see Fig. 7 and Fig. 8) in the Right Arc which is absent in Left arc. This suggests that this tidal arm is older and has hosted star formation over longer timescales, and may have already exhausted its molecular gas reservoir. Furthermore, recent ALMA observations confirm the absence of CO in most of the Right Arc and SDR at high spatial resolution (P. N. Appleton, priv. comm.), consistent with our ACA-based results. The only exception is the USDR, which shows considerable CO emission in the ALMA data. However, the overall consistency indicates that our findings are not strongly limited by ACA’s spatial resolution but instead reflect a genuine lack of CO in these regions.

It has been shown that the star formation in SQ-A probably started before the main shock event ($<$ 10 Myr ago), and this probably explains the large number of old clusters ($\tau > 10^8$ yr) around the SQ-A region \citep{guillard2012, konstant2014}.
However, it should be noted that clusters older than $10^8$ yr are also present in LeftArc which hosts the main shock and the spiral southern arm of 7318b, although more scattered. This may indicate the possibility of older shock events and subsequent cooling of the shocked gas that occurred hundreds of Myr ago. 
As mentioned before, \citet{Moles1997} concluded that NGC7320C collided with the group a few times 10$^8$ years ago.
It is tempting to say that the region might have experienced multiple episodes of shock-induced star formation over an extended period.
However, it is important to keep in mind that the three-dimensional distribution of clusters in regions like the Left arc remains uncertain. 
An NIR IFU study of some of these clusters which may be less affected by dust extinction effects will be able to provide important clues in this regard.

\subsubsection{Comparison with previous SFR estimates in the literature}
The estimated total SFR from our derived SCC properties, as discussed in \ref{subsec:spatial_scc}, turns out to be 0.126 \msun/yr. 
Note that the region considered in our study does not include several star-forming regions such as SQB and also miss out on many clusters in regions like SQA due to strong extended emission in UV/optical bands.
We find that our estimate is still significantly lower from the SFR estimates calculated using other methods such as using H$\alpha$ emission in previous works.
Using the H$\alpha$ emission coming from $\sim 30$ \hii\ regions of young star clusters in LeftArc, SQ-A and SDR regions, \citet{konstant2014} estimated the total SFR in the region to be 0.084 \msun/yr.
But these regions were biased towards the most luminous (massive) \hii\ regions and hence may not represent a complete census of young star clusters as we try to achieve with our HST selected sample.
However, \citet{puertas2021} covered almost the entirety of SQ using SITELLE (an imaging Fourier transform spectrometer, attached to the Canada-France-Hawaii Telescope) and found a higher SFR of $\sim 3.1$ \msun/yr from $\sim 100$ H{\sc ii} regions over the whole region. They found SQ-A contributing almost 28$\%$ of the total SFR.
Although this work only considered SFR from regions that showed clear H{\sc ii} region signatures spectroscopically, it is still possible that some fraction of the H$_{\alpha}$ emission within their apertures could be contaminated by shocked gas. This can lead to an overestimation of SFR in such studies.

In other works, \citet{Cluver2010} estimated the star formation along the main filament along the Left arc using a combination of H$_{\alpha}$ and 24 micron continuum measurements and found a total SFR of $< 0.05$ \msun/yr, and using the 7.7 $\mu$m PAH emission, found a similarly low value of 0.08 \msun/yr. 
Interestingly, this is roughly consistent with the total estimated SFR of $\sim0.06$ \msun/yr for the Left arc in our study.
Additionally, \citet{guillard2012} found that the low 7.7 $\mu$m PAH/CO surface density found in the ridge was evidence for very low star formation efficiency suggesting overall star formation suppression in the main ridge. These low estimates may be influenced by possible PAH destruction in the shock.  A higher star formation rate across the main filament (1 \msun/yr) was found by \citet{natale2010} using a combination of UV and far-IR measurements. This may be an upper limit because \citet{guillard2022} showed that a significant amount of the UV emission is dominated by broad powerful Ly-$\alpha$ emission from shocked gas. Also, \citet{natale2010} estimated the total SFR across the whole group to be 7.5 \msun/yr, again assuming the entire UV emission is generated by star formation.

The higher SFR values reported in previous studies may indicate that our analysis is missing a fraction of star clusters, particularly young massive clusters (YMCs), due to both our strict SCC selection criteria and the limitations of UV/optical-based detection methods. 
Notably, the possible presence of even a few tens of YMCs could significantly raise our SFR estimates, bringing them closer to those reported by \citet{puertas2021}.
To assess the extent to which our sample selection (as described in Section~\ref{subsec:sel_criteria}) influences the derived SFRs, we relaxed the selection criteria—retaining only the color cuts—and constructed a broader sample of detected point sources. We then re-ran {\sc Cigale} on this expanded sample and found that the estimated SFR increased substantially to $\sim$1 \msun/yr, compared to $\sim0.13$ \msun/yr from our high-confidence SCC sample.
This significant difference suggests that our conservative selection aimed at minimizing contamination and background sources can lead to an underestimation of the SFR. These findings highlight the need for future spectroscopic observations with higher spatial resolution to improve the completeness of the cluster population in Stephan’s Quintet.

\section{Conclusions}
\label{sec:conclusions}

In this work, we aim to derive physical properties of star clusters in the famous compact galaxy group known as Stephan's Quintet (SQ).
For this, we use a catalog of 1,588 high confidence star cluster candidates (SCC) created from HST images after applying several selection criteria to a list of detected point sources found in the SQ field.
We obtain the SCC magnitudes in HST \U-\B-\Vm-\V-\I\ bands in addition to NIRCam F090W, F150W, F200W, F277W and F356W bands in the near-infrared from the recent JWST observations.
Taking advantage of this large photometric baseline, we carry out SED fitting using {\sc Cigale} to derive key SCC properties such as cluster age ($\tau$), cluster mass (m) and dust extinction ($A_V$).
We performed {\sc Cigale} runs combining both HST and JWST photometry and HST photometry alone to understand the effects of NIR photometry in SED fitting results.
Our main results are as follows :

\begin{itemize}
    \item We find that both HST+JWST and HST only runs are largely consistent in terms of the derived SCC properties. They show some general trends as follows : (1) The age distribution shows two prominent peaks, one around $\sim 5$ Myr and the second one at $\sim 2 \times 10^8$ yr. The former corresponds to an estimated timescale of the recent collision of NGC 7318b with the group and the latter with that of an older interaction with 7320c.
    (2) A majority of SCCs ($\sim50\%$) show little extinction ($A_V < 0.5$) while a considerable fraction of young SCCs ($\sim25\%$) show moderate extinction values ( $A_V > 1.0$).
    (3) The cluster mass shows a distribution skewed towards the low-mass end (m $ < 10^{4.5}$\msun) with a few older clusters with higher masses (m $> 10^{5}$\msun).

    \item The inclusion of NIR photometry in addition to UV/optical in SED fitting routines in {\sc Cigale} leads to considerable changes in the derived parameters of many SCCs and might help in breaking the difficult problem of age-extinction degeneracy. We find 121 ($8\%$) SCCs shifted from older ages ($\tau > 10^8$ yr) to young ages ($\tau < 10^7$ yr) with a corresponding increase in $A_V$ after including the JWST NIRCam filters.

    \item From cluster age versus mass diagram of SCCs in different regions of SQ, we observe a complete lack of young ($\tau < 10^7$ yr) massive (m $>10^5$\msun) clusters in our sample. It is possible that our sample selection criteria, which require detection across UV/optical images, miss out on young massive clusters that may be still embedded in dust. However, the unusual nature (shock-triggered) and location (external to a galaxy disk) of many of the young star clusters may inhibit the formation of massive clusters.

    \item Consistent with previous results, the spatial maps show that most of the young clusters are tightly clustered along the shock regions in the left and right arc regions around the 7318a/b merger (Fig~\ref{fig:age_sq}). This is a strong indication that the recent star formation is triggered by the 7318a/b galaxy collision. Additionally, many young clusters are present in SQ-A and the Southern Debris regions.  The older clusters ($\tau > 10^8$ yr) are more uniformly distributed around the group. 

    \item The ACA CO(2-1) emission map (Fig~\ref{fig:molecular_map}) shows young clusters in the Left arc and SQ-A are mostly clustered around regions of peak CO emission whereas most of the Right arc shows a complete lack of CO emission despite the large number of young clusters in the region.

    \item We estimate the current total SFR within clusters to be $\sim$0.13~\msun/yr, which is significantly lower than several SFR estimates reported for SQ in the literature. This discrepancy could be due to the apparent absence of young massive clusters in our sample that are potentially undetected in UV/optical images. Additionally, our selection methods, that are based on point source detection, may exclude SCCs in regions of strong extended emission such as SQ-A. 
\end{itemize}

Although NIR photometry improves the SED fits in our study, we also find that this can introduce several limitations, such as (i) potential contamination from red supergiants (RSGs) and pulsating AGB stars which may be especially important in the absence of massive clusters in the field, (ii) projection effects, and (iii) assumptions inherent to the "energy balance" principle employed in {\sc Cigale} (see Subsection~\ref{subsec:discuss1_NIR} for detailed discussions).  
We aim to address these challenges and further refine our estimates on SCC properties in future analyses.

Further, we plan to continue a detailed investigation into the nature of SCCs in Stephan's Quintet using NIRCam and MIRI images. The NIRCam-only detections (Aromal et al., in preparation) may reveal young massive clusters (YMCs) in SQ that are still in their embedded phase and hence improve the SFR estimates as well.
In parallel, we have recently acquired high-resolution ALMA data (PI : P. N. Appleton) covering large regions of SQ. Combined with our SCC sample, this will provide valuable insights into the various processes influencing star formation and star formation efficiency at smaller scales. 
Additionally, we aim to improve dust emission models in the NIR to better disentangle contributions from red supergiants (RSGs), asymptotic giant branch (AGB) stars, and dust emission from the clusters itself.
This will further improve breaking age-extinction degeneracy using NIR filters in the future studies.

\section*{Acknowledgements}

This work is supported by the Canadian Space Agency, the Natural Science and Engineering Research Council (NSERC) RGPIN-2021- 04157 and a Western Research Leadership Chair Award.
P. N. Appleton acknowledges support under NASA Guest Observer grant JWST-GO-03445 001-A.
UL acknowledges support by the research grant  PID2023-150178NB-I00, financed by MCIU/AEI/10.13039/501100011033  and from the Junta de Andaluc\'{i}a (Spain) grant  FQM108.
The National Radio Astronomy Observatory is a facility of the National Science Foundation operated under cooperative agreement by Associated Universities, Inc.

This research made use of Photutils, an Astropy package for detection and photometry of astronomical sources \citep{larry_bradley_2024}.

\begin{contribution}


\end{contribution}

%
\facilities{HST(STIS), JWST(STIS)}

\software{astropy \citep{astropy2022},  
          {\sc Cigale} \citep{cigale2019}
          }





\bibliography{bib_scc, hcgroupies}{}
\bibliographystyle{aasjournalv7}



\end{document}